\documentclass[preprint,12pt,authoryear]{elsarticle}

\usepackage{amssymb}

\usepackage[colorlinks=true,linkcolor=blue, citecolor=blue]{hyperref}

\usepackage{natbib}
\usepackage{amsthm}
\usepackage{amsmath}
\usepackage{mathrsfs}
\usepackage{graphicx}
\usepackage{epstopdf}
\usepackage{float}
\usepackage{caption}
\usepackage{bm}
\usepackage{amssymb} 
\usepackage{extarrows} 
\usepackage{bbm}
\usepackage{mathrsfs}
\usepackage{cleveref}
\usepackage{soul}
\usepackage{tabu} 
\usepackage{longtable}
\usepackage[margin=2cm]{geometry}
\usepackage{xcolor} 

\usepackage[caption=false]{subfig}

\usepackage[version=4]{mhchem}
\usepackage{siunitx}
\usepackage{array}

\journal{Journal of the Mechanics and Physics of Solids}

\makeatletter
\def\@author#1{\g@addto@macro\elsauthors{\normalsize%
    \def\baselinestretch{1}%
    \upshape\authorsep#1\unskip\textsuperscript{%
      \ifx\@fnmark\@empty\else\unskip\sep\@fnmark\let\sep=,\fi
      \ifx\@corref\@empty\else\unskip\sep\@corref\let\sep=,\fi
      }%
    \def\authorsep{\unskip,\space}%
    \global\let\@fnmark\@empty
    \global\let\@corref\@empty  
    \global\let\sep\@empty}%
    \@eadauthor={#1}
}
\makeatother

\newcommand{\Li}{\mathrm{Li}}
\newcommand{\Lat}{\mathrm{L}}

\newcommand{\Gibbs}{\mathcal{G}}
\newcommand{\bulk}{\mathrm{Bulk}}
\newcommand{\inter}{\mathrm{Interface}}
\newcommand{\dev}{\mathrm{dev}}

\newcommand{\m}{\mathrm{m}}
\newcommand{\vac}{\mathrm{v}}

\usepackage{setspace}
\renewcommand{\baselinestretch}{1.5}

\begin{document}

\begin{frontmatter}



\title{A phase field electro-chemo-mechanical formulation for predicting void evolution at the Li-electrolyte interface in all-solid-state batteries}


\author{Ying Zhao\corref{cor1}\fnref{Tongji}}
\ead{19531@tongji.edu.cn}

\author{Runzi Wang \fnref{IC}}

\author{Emilio Mart\'{\i}nez-Pa\~neda\corref{cor1}\fnref{IC}}
\ead{e.martinez-paneda@imperial.ac.uk}

\address[Tongji]{School of Aerospace Engineering and Applied Mechanics, Tongji University, Shanghai 200092, China}

\address[IC]{Department of Civil and Environmental Engineering, Imperial College London, London SW7 2AZ, UK}

\cortext[cor1]{Corresponding authors.}

\begin{abstract}
We present a mechanistic theory for predicting void evolution in the Li metal electrode during the charge and discharge of all-solid-state battery cells. A phase field formulation is developed to model vacancy annihilation and nucleation, and to enable the tracking of the void-Li metal interface. This is coupled with a viscoplastic description of Li deformation, to capture creep effects, and a mass transfer formulation accounting for substitutional (bulk and surface) Li diffusion and current-driven flux. Moreover, we incorporate the interaction between the electrode and the solid electrolyte, resolving the coupled electro-chemical-mechanical problem in both domains. This enables predicting the electrolyte current distribution and thus the emergence of local current `hot spots', which act as precursors for dendrite formation and cell death. The theoretical framework is numerically implemented, and single and multiple void case studies are carried out to predict the evolution of voids and current hot spots as a function of the applied pressure, material properties and charge (magnitude and cycle history). For both plating and stripping, insight is gained into the interplay between bulk diffusion, Li dissolution and deposition, creep, and the nucleation and annihilation of vacancies. The model is shown to capture the main experimental observations, including not only key features of electrolyte current and void morphology but also the sensitivity to the applied current, the role of pressure in increasing the electrode-electrolyte contact area, and the dominance of creep over vacancy diffusion.\\ 


\end{abstract}


\end{frontmatter}


\section{Introduction}
\label{Sec:Intro}

\begin{figure}[H]
\includegraphics[width=\linewidth]{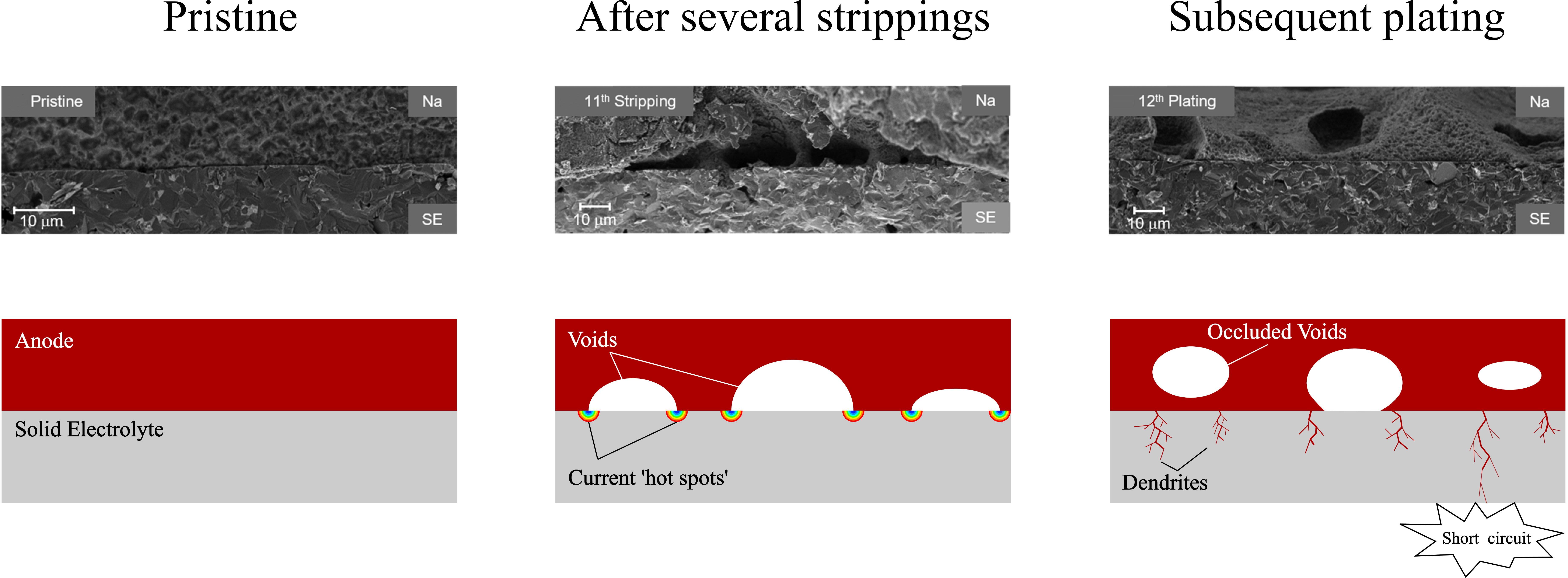}
\caption{Interplay between voiding and dendrite formation: SEM images and sketches of the process of voiding and dendrite formation during plating and stripping cycles. The presence of voids generates local current `hot spots' at the locations where the void meets the Li anode and the solid electrolyte. Li deposition will be exacerbated in these regions and lead to the nucleation of dendrites, which grow until causing the short circuit of the cell. The SEM images are based on the work by \citet{SpencerJolly2020} using Na anodes but are representative of what is observed in other all-solid-state battery systems, including Li-based.}\label{fig:introduction}
\end{figure}

All-solid-state batteries are arguably the most promising development in energy storage technology, allowing for significant improvements in both energy and power densities, relative to traditional Li-ion batteries using organic liquid electrolytes \citep{Pasta2020}. In particular, all-solid-state batteries employing lithium (Li) metal anodes constitute a very promising candidate given their superior electrochemical performance among solid-electrolyte systems \citep{Wu2021a}. However, void and dendrite formation at the interface between the Li-metal anode and the solid electrolyte are hindering the viability of all-solid-state Li-metal batteries \citep{Janek2016}. Lithium dendrites can penetrate the separator and short circuit the battery cell, while voids reduce the contact area between the electrode and the solid electrolyte, significantly increasing the cell's internal resistance. Both failure phenomena are interfacial instabilities that are likely consequences of unevenly distributed current at the electrode-electrolyte interface. How and when the current inhomogeneity can result in dendrite formation is yet to be understood \citep{Pang2021}. Moreover, dendrite development and voiding are strongly coupled processes - see Fig. \ref{fig:introduction}.\\ 

During discharge, Li ions move from the Li anode to the cathode and \emph{stripping} takes place; metallic lithium undergoes anodic dissolution and this leads to the nucleation and growth of voids in the Li metal anode at its interface with the solid electrolyte. As a result, local regions of high current density (`hot spots') emerge at the electrolyte, in the vicinity of the junction with the Li anode and the void. During charging, metallic lithium is deposited in a process termed \emph{plating}, which results in a reduction of the voids' size and, occasionally, in the occlusion of voids formed on the previous stripping cycle (see Fig. \ref{fig:introduction}). More importantly, Li plating is exacerbated at hot spots, leading to the nucleation and subsequent propagation of dendrites, needle-like structures between electrodes that short circuit the battery cell. This process has been recently shown to be very sensitive to the plating and stripping currents \citep{Kasemchainan2019}. A critical plating current density exists, above which dendrites nucleate, and voids form when the stripping current density exceeds the rate at which Li is replenished at the surface. There is a need to predict the evolution of voids during multiple plating/stripping cycles and map the conditions that lead to voiding, dendrite formation and cell death.\\

Recently, it has been suggested that voiding and subsequent dendrite formation can be suppressed by applying a constant pressure to the free surface of the Li metal anode \citep{Sakamoto2019}. The aim is to maintain a void-free electrode-electrolyte interface and significantly improve the interfacial contact area due to creep deformation. Recent experimental studies have shown that voiding can be hindered if the applied pressure is sufficiently high such that the rate at which Li metal flows to the interface is larger than the Li stripping rate \citep{Krauskopf2019,Wang2019a,Kasemchainan2019}. Nonetheless, it is still not clear how the interplay between pressure, Li creep, diffusion, voiding, and current density governs electrochemical instabilities in an all-solid-state battery cell. Theoretical and computational electro-chemo-mechanical formulations are needed to capture the behaviour, over multiple charge/discharge cycles, of the interfacial phenomena governing the stability of all-solid-state batteries.\\ 

Monroe and Newman were among the first to propose a mechanistic rationale for dendrite formation and an accompanying model \citep{Monroe2003,Monroe2004,Monroe2005}. Their modelling efforts suggested that dendrites could be suppressed by increasing the stiffness of the solid electrolyte beyond twice that of lithium. However, experimental endeavours using ceramic electrolytes have proved otherwise \citep{Schmidt2016,Ren2015}. Stiff ceramic electrolytes contain defects such as grain boundaries, pores and pre-existing flaws that facilitate local Li deposition and trigger the growth of cracks and dendrites \citep{Raj2017,Porz2017}. Increasing electrolyte stiffness can also favour interface delamination, as shown computationally by \citet{Rezaei2021}. Moreover, \citet{Porz2017} revealed that Li penetration occurred even when single-crystal solid electrolytes were used. The stability of the electrode-electrolyte interface could be improved by the application of an elastic pre-stress, as shown theoretically by \citet{Natsiavas2016}, or by the use of an artificial Solid-Electrolyte-Interphase (SEI) with a sufficiently large stiffness, as shown by \citet{Liu2019a}. Studies have also highlighted how Li penetration is favoured by the viscoplastic and corrosive nature of metallic lithium \citep{Ma2016a,Narayan2018,Wang2020b}, emphasising the importance of an adequate characterisation of its mechanical behaviour \citep{Anand2019a}. Several simplified electro-chemo-mechanical models have been recently developed to gain insight into the conditions triggering dendrite formation. \citet{Klinsmann2019} used an analytical model to investigate the ability of a crack to propagate across the solid electrolyte. They concluded that a sufficiently high pressure could block the redox reaction and suppress dendrite growth. \citet{Shishvan2020c,Shishvan2020d} idealised dendrite growth as the climbing of an edge dislocation and used this analogy to predict critical current densities for the propagation of dendrites. \citet{Narayan2020} presented a modelling strategy to simulate plating and stripping at the anode-electrolyte interface and studied the role of interface impurities and surface defects. \citet{Roy2021} simulated interfacial flux in the presence of a debonded patch and concluded that standard Butler–Volmer kinetics are inadequate to explain experimental observations. However, these models mainly focus on dendrite formation and cannot predict the dynamic response of the electrode-electrolyte interface during cycling, nor the nucleation and evolution of voids in the Li metal anode. Predicting voiding and the associated localisation of current requires resolving the coupled diffusion, electrical and mechanical problems, and computationally tracking the evolution and merging/division of voids of arbitrary shape - a formidable task.\\ 

In this work, we present a new electro-chemo-mechanical formulation for all-solid-state batteries which is capable of predicting void evolution in the Li metal anode and the electrolyte current density distribution. Our theoretical and computational framework is developed by building upon the phase field method, which enables capturing relevant morphological changes in the void-electrode interface such as shape variations of arbitrary complexity (including occlusion) and the interaction between multiple voids (e.g., void coalescence). In addition, our model accounts for: (i) the nucleation and annihilation of Li metal lattice sites, (ii) the diffusion of Li in the electrode, (iii) the viscoplastic and creep behaviour of the Li metal anode, and (iv) the interaction between the electrode and the solid electrolyte. The theoretical model presented is numerically implemented and simulations are conducted over multiple plating/stripping cycles to gain insight into how voids and hot spots for dendrite formation evolve as a function of the applied pressure, material properties, and charge.


\section{Theory}
\label{Sec:Theory}

\subsection{Preliminaries}

\begin{figure}[H]
    \centering
    \makebox[\textwidth]{
    \includegraphics[width=1.1\linewidth]{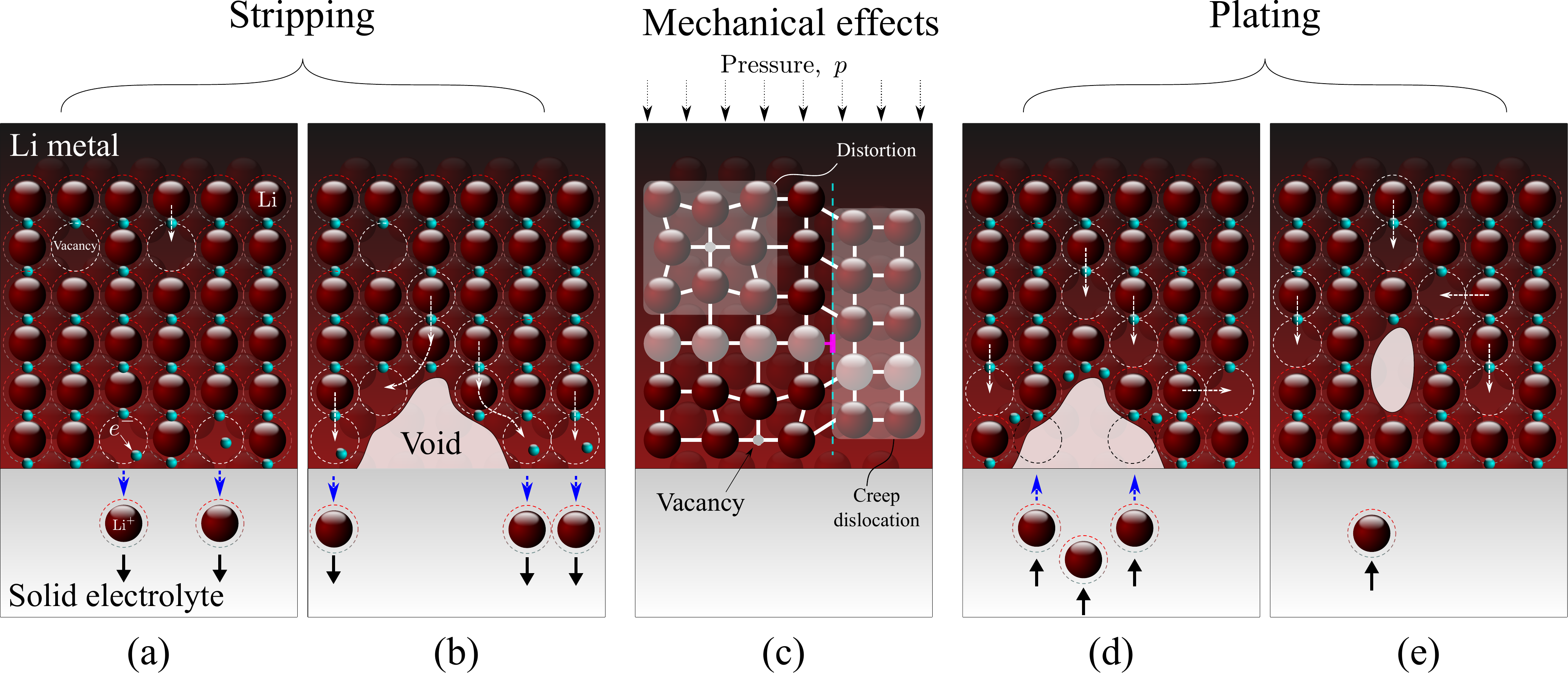}}
    \caption{Schematic of the physical mechanisms governing voiding at the anode-electrolyte interface of all-solid-state batteries: (a) initial stripping stage, with vacancy diffusion and Li dissolution; (b) advanced stripping stage, with voiding, Li dissolution and both adatom and vacancy diffusion; (c) mechanical interactions relevant to both stripping and plating - vacancy diffusion distorts the lattice while creep due to an applied pressure closes the voids; (d) initial plating stage, with Li deposition reversing void growth; and (e) advanced plating stage, with void occlusion being observed due to further Li deposition.}
    \label{fig:SketchPhenomena}
\end{figure}

Our theory aims at encapsulating the main physical mechanisms driving void nucleation and evolution during plating and stripping (see, e.g., \citealp{Krauskopf2019}). The phenomena at play are sketched in Fig. \ref{fig:SketchPhenomena}.

Consider first the case of stripping. At the beginning of the process (Fig. \ref{fig:SketchPhenomena}a), two mechanisms are relevant: (i) substitutional diffusion within the bulk of the Li metal anode, where Li atoms diffuse to adjacent vacant sites, and (ii) dissolution of Li atoms at the anode-electrolyte interface, with Li ions moving to vacant or interstitial sites in the electrolyte and leaving behind in the anode an electron and a vacancy. If the current density is sufficiently large, the stripping rate of metal ions at the surface will exceed that of bulk vacancy diffusion, leading to void nucleation and contact loss (Fig. \ref{fig:SketchPhenomena}b). Specifically, voids form because vacancies condensate \citep{Cuitino1996} or annihilate at available sinks such as dislocations, grain boundaries and free surfaces. Void growth is further enhanced by adatom diffusion, as the transport of atoms along the void surfaces is faster than bulk vacancy diffusion \citep{Jackle2018}. Voiding can be minimised by the application of mechanical pressure, as shown in Fig. \ref{fig:SketchPhenomena}c. Due to the viscoplastic nature of metallic lithium, plastic deformation and creep become relevant at sufficiently high pressures and act to close the voids and reduce contact loss. As depicted in Fig. \ref{fig:SketchPhenomena}c, another interaction with mechanics is the contraction and expansion of the Li atomic lattice in the anode, as a result of vacancy diffusion. Finally, consider the plating process (Figs. \ref{fig:SketchPhenomena}d and \ref{fig:SketchPhenomena}e). Li ions move through the electrolyte into the anode, forming a Li atom in a surface site through vacancy nucleation and the interaction with an electron present in the anode. If the deposition current density is sufficiently high, void growth will be reversed and surface contact improved, with voids often becoming occluded from the interface.\\ 

Thus, voiding is governed by substitutional bulk and surface diffusion, lattice distortion, Li dissolution and deposition, nucleation and annihilation of vacancies, and creep deformation. We capture these phenomena by means of a coupled diffusion-deformation-phase field theory. The phase field paradigm is exploited to describe the evolution of the void-Li metal anode interface. Thus, the phase field order parameter ($\xi$) takes the values of 0 and 1 at the void and Li metal phases, respectively, and is defined to evolve as dictated by vacancy annihilation and nucleation. The diffusion problem takes as primary kinematic variable the occupancy of Li sites in the anode ($\theta_m$) and as boundary condition a current-dependent flux that incorporates the role of Li dissolution and deposition. Moreover, both bulk and surface diffusion are captured in its governing equation through the interaction with the mechanical and phase field problems. Finally, as appropriate for metallic lithium, the mechanical description is characterised by a viscoplastic constitutive response, capturing creep effects. Also, the role of lattice distortion is incorporated \textit{via} chemical strains and the balance of linear momentum is coupled to the phase field to account for the presence of voids. These elements of our theory are defined below in a thermodynamically-consistent manner. Moreover, the framework is extended to solve for the electric potential in the electrolyte, yielding a coupled electro-chemo-mechanical formulation able to predict the occurrence of local current hot-spots, which result from the voiding process and lead to the formation of dendrites.

\subsection{Gibbs Free energy and chemical potential of the lithium metal electrode}

\begin{figure}[H]
\centering
\includegraphics[width=0.8\linewidth]{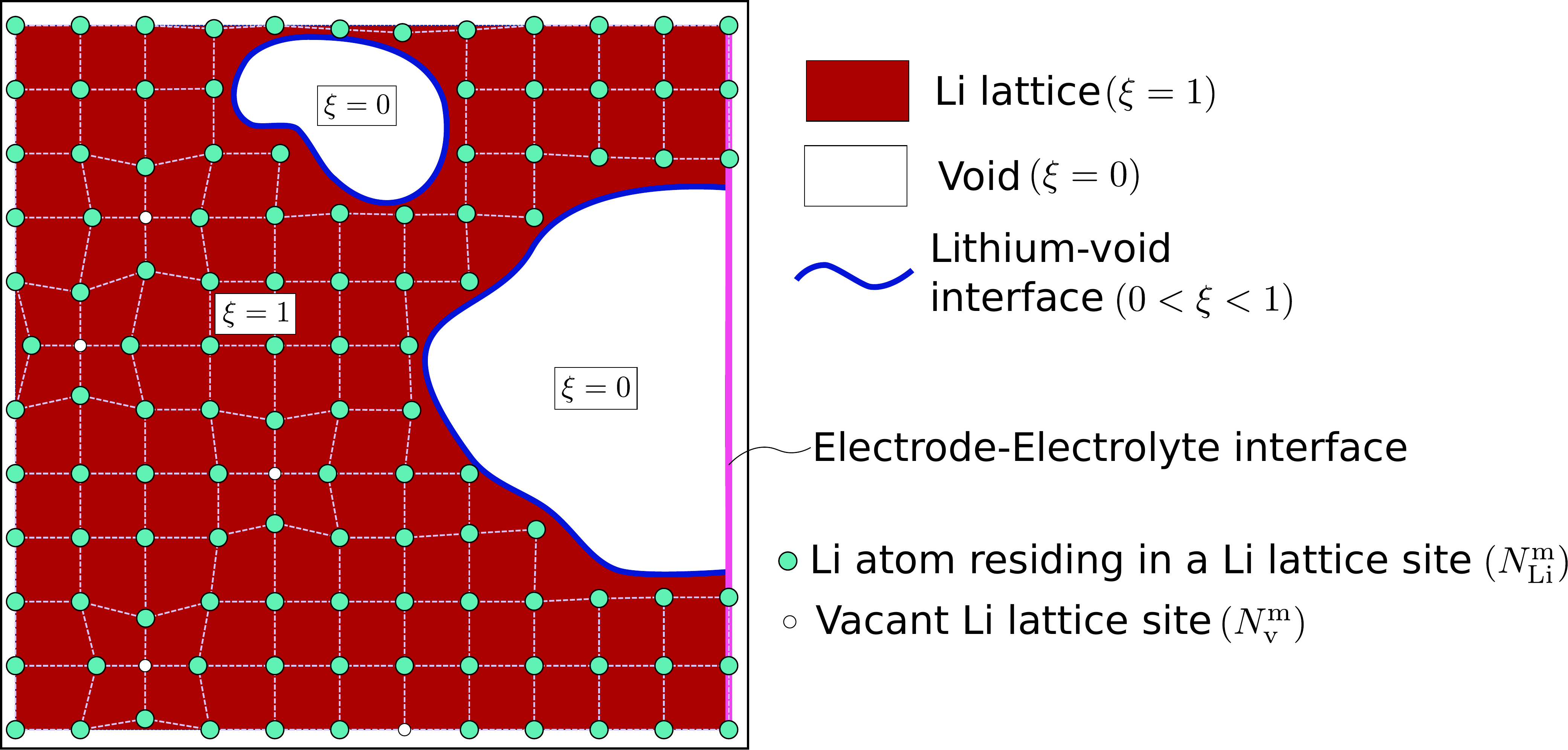}
\caption{Illustration of the sites in a volume $V$ of lithium metal. The shaded area denotes the network of lithium lattice sites and takes over a volume $V^\m$. Thus, the voids occupy a volume of $V^\vac = V-V^\m$. In $V^\m$, there are $N_\Li^\m$ lattice sites occupied by lithium atoms, leaving $N_\vac^\m$ lattice sites vacant.}\label{fig:Illustration_sites}
\end{figure}

Consider a lithium metal electrode with a volume $V$, where the Li lattice sites can be annihilated into and nucleated from voids, as shown in Fig. \ref{fig:Illustration_sites}. At some instant of time $t$, there are $N_\Lat^\m$ moles of lattice sites in the Li metal network, where $N_\Li^\m$ moles of Li atoms reside, leaving $N_\vac^\m$ moles of lattice sites vacant. Throughout this manuscript, we use ``$\m$'' as a superscript or a subscript to denote variables related to the metallic electrode.

Let $\Omega_\Li$ and $\Omega_\vac$ respectively denote the molar volume of Li and vacant sites. Then, the total volume of the Li metal network equals
\begin{align}
V^\m = N_\Li^\m\Omega_\Li + N_\vac^\m\Omega_\vac.\label{eq:vm}
\end{align}

\noindent The remaining volume, denoted by $V^\vac$, corresponds to the voids. The total volume of the anode is thus $V = V^\m + V^\vac$, and is kept constant. We emphasise that voids refer to regions where vacancies have been condensated or annihilated; unlike vacancies, voids cannot store strain energy. In the absence of external loading, vacancies are in equilibrium with voids. We proceed to define the molar concentration of Li atoms $c_\Li^\m$, Li vacant sites $c_\vac^\m$, and Li lattice sites $c_\Lat^\m$ as
\begin{align}
c_\Li^\m = \lim_{\mathrm{d}V^\m\to0}\frac{\mathrm{d}N_\Li^\m}{\mathrm{d}V^\m}, \quad c_\vac^\m =  \lim_{\mathrm{d}V^\m\to0}\frac{\mathrm{d}N_\vac^\m}{\mathrm{d}V^\m},\quad c_\Lat^\m =  \lim_{\mathrm{d}V^\m\to0}\frac{\mathrm{d}N_\Lat^\m}{\mathrm{d}V^\m}.\label{eq:concentrations}
\end{align}
Let $\theta_\m = c_\Li^\m/c_\Lat^\m$ denote the Li occupancy, while $\theta_\vac=c_\vac^\m/c_\Lat^\m$ gives the vacancy occupancy in the Li metal network. It follows that $\theta_\vac = 1-\theta_\m$. In the Li metal anode, Li is both the solute and the solvent, implying that there is no theoretical upper bound on the Li concentration when $V_\m$ reduces to 0 \citep{Lin2017}. In such a scenario (e.g., stripping), the present phase field model will predict a transformation of the Li metal phase into void phase, and all concentrations will automatically become equal to 0 (as elaborated below).\\

For a pressure $p_\m$ and an elastic strain energy density $\psi_e$, the Gibbs free energy of the electrode reads \citep{Shishvan2020c}
\begin{align}
    \Gibbs^\m = N_\Li^\m\mu_\m^0 + N_\vac^\m h_\vac + N_\Lat^\m RT\left[\theta_\m\ln{\theta_\m} + \left(1-\theta_\m\right)\ln{\left(1-\theta_\m\right)}\right] +V^\m\left(\psi_e+p_\m\right)\label{eq:Gibbs_free_energy_Li_metal_old}
\end{align}
where $R$ is the gas constant, $T$ is the absolute temperature, $\mu_\m^0$ is the reference molar enthalpy of Li atoms at zero pressure and $h_\vac$ is the molar enthalpy of the formation of vacant sites in Li ($\approx 50$ kJ/mol, \citealp{Shishvan2020c}). Assuming that the void is subjected to the same pressure as the lithium metal, the Gibbs free energy of the void is
\begin{align}
\Gibbs^\vac = \left(V-V^\m\right)p_\m.
\end{align}
The total Gibbs free energy, discarding the interfacial energy between the void and the lithium metal, is thus
\begin{align}
\Gibbs = N_\Li^\m\mu_\m^0 + N_\vac^\m h_\vac + N_\Lat^\m RT\left[\theta_\m\ln{\theta_\m} + \left(1-\theta_\m\right)\ln{\left(1-\theta_\m\right)}\right] +V^\m\psi_e  +V p_\m
\end{align}
The chemical potentials of Li atoms and lattice sites can be respectively derived as
\begin{align}
    &\mu_\Li^\m = \left.\frac{\delta\Gibbs^\m}{\delta N_\Li^\m}\right\vert_{N_\Lat^\m} = \left(\mu_\m^0-h_\vac\right) + RT\ln{\frac{\theta_\m}{1-\theta_\m}} + \left(\Omega_\Li-\Omega_\vac\right)\left(\psi_e+p_\m\right),\label{eq:mu_Li}\\
    &\mu_\Lat^\m = \left.\frac{\delta\Gibbs^\m}{\delta N_\Lat^\m}\right\vert_{N_\Li^\m} = h_\vac + RT\ln{\left(1-\theta_\m\right)} + \Omega_\vac\left(\psi_e+p_\m \right).\label{eq:mu_L}
\end{align}
We shall now explore the equilibrium chemical potentials $\mu_\m^0$ and $h_\vac$. In equilibrium, diffusion of lithium atoms is precluded and thus the chemical potential is constant everywhere, viz. $\mu_\Li^\m = \mu_\Lat^\m$. Moreover, Li vacant sites are not annihilated or nucleated under equilibrium conditions, indicating that $\mu_\Lat^\m$ should vanish. Thus, $\mu_\Li^\m=\mu_\Lat^\m=0$. If we define $\theta_\m^0$ as the equilibrium lithium occupancy in the absence of mechanical pressure and elastic straining, it naturally follows from Eqs. (\ref{eq:mu_Li}) and (\ref{eq:mu_L}) that
\begin{equation}
h_\vac + RT\ln\left(1-\theta_\m^0\right) = 0 \, , \,\,\,\,\,\,\,\, \text{and} \,\,\,\,\,\,\,\, \mu_\m^0  + RT\ln\theta_\m^0 = 0 \, .
\end{equation}
Accordingly, we can derive $\theta_\m^0$ and $\mu_\m^0$ as
\begin{equation}
\theta_\m^0 = 1-\exp\left(-\frac{h_\vac}{RT}\right) \, ,\,\,\,\,\,\,\, \text{and} \,\,\,\,\,\,\,\, \mu_\m^0 = -RT\ln\left[1-\exp\left(-\frac{h_\vac}{RT}\right)\right] \, .
\end{equation}

Thus, the total Gibbs free energy of the lithium electrode is
\begin{align}
\Gibbs^\bulk = N_\Lat^\m RT\left[\theta_\m\ln{\frac{\theta_\m}{\theta_\m^0}}+ \left(1-\theta_\m\right)\ln{\frac{1-\theta_\m}{1-\theta_\m^0}}\right] +V^\m\psi_e+V p_\m \, ,
\end{align}

\noindent where the superscript ``Bulk" is used to emphasise the absence of a Li metal-void interfacial energy term. We can also obtain the Helmholtz free energy from $\Psi^\bulk = \Gibbs^\bulk-V p_\m$, yielding
\begin{align}
\Psi^\bulk = N_\Lat^\m RT\left[\theta_\m\ln{\frac{\theta_\m}{\theta_\m^0}}+ \left(1-\theta_\m\right)\ln{\frac{1-\theta_\m}{1-\theta_\m^0}}\right] +V^\m\psi_e \, ,
\end{align}

\noindent which, recalling Eq. (\ref{eq:concentrations}), can be reformulated as
\begin{align}
\Psi^\bulk =  \int_{V^\m} \left\{ c_\Lat^\m RT\left[\theta_\m\ln{\frac{\theta_\m}{\theta_\m^0}}+ \left(1-\theta_\m\right)\ln{\frac{1-\theta_\m}{1-\theta_\m^0}}\right] + \psi_e \right\} \mathrm{d}V \, , \label{eq:Helmholtz_li_metal}
\end{align}

\subsection{Phase field formulation of the Helmholtz free energy of a lithium electrode with voids}

Let us now incorporate into the model the presence of an evolving void-anode interface. To achieve this, we exploit the phase field paradigm, which has been successfully applied to a wide range of interfacial problems, including corrosion \citep{JMPS2021,Ansari2021}, microstructural evolution \citep{Li2018c,Ask2018}, fracture \citep{Bourdin2000,JMPS2020}, and dendrite growth in liquid electrolytes \citep{Chen2015a,Chen2021a}. Thus, an order parameter $\xi= \xi(\mathbf{x},t)$ is introduced, which is a time-dependent field variable defined in the volume $V$. As shown in Fig. \ref{fig:Illustration_sites}, it bears the value of 1 in the lattice and 0 in the void, varying smoothly in-between. Further, we introduce $\xi$-dependent variables: lattice concentration $c_\Lat^\xi$, lithium occupancy $\theta_\m^\xi$ and elastic strain energy density $\psi_e^\xi$ as
\begin{align}
&c_\Lat^\xi = c_\Lat^\m,\quad \theta_\m^\xi = \theta_\m,\quad \psi_e^\xi = \psi_e &&\text{ in }V^\m (\xi = 1), \text{ and}\label{eq:ctp_xi1}\\
&c_\Lat^\xi = 0,\quad \theta_\m^\xi = \theta_\m^\mathrm{const},\quad \psi_e^\xi = 0 &&\text{ in }V^\vac (\xi = 0).\label{eq:ctp_xi0}
\end{align}
The lattice concentration $c_\Lat^\m$ is taken as a constant, with the average molar volume of the lattice $\Omega_\Lat$ as reciprocal. In the void, there are no lattice sites ($c_\Lat^\xi=0$) and thus the model is, from a theoretical viewpoint, insensitive to the choice of $\theta_\m^\xi$ in $V^\vac (\xi = 0)$. However, note that numerical convergence is facilitated by choices of $\theta_\m^\mathrm{const}$ that are close to the magnitude of the lithium occupancy in the lattice $\theta_\m$. Since $\theta_\m^\xi \equiv \theta_\m$ for $0 < \xi \leq 1$, we proceed to drop the superscript $\xi$ for the lithium occupancy. Contrarily, $c_\Lat^\xi$ and $\psi_e^\xi$ are interpolated along the lithium-void interface by means of an interpolation function $h(\xi)$, such that
\begin{align}
c_\Lat^\xi = h(\xi) c_\Lat^\m \, ,  \quad \quad \text{and} \quad \quad \psi_e^\xi = h(\xi) \psi_e \, . \label{eq:inter_cp}
\end{align}

This interpolation function must satisfy $h(0)=0$ and $h(1)=1$, such that Eqs. (\ref{eq:ctp_xi1}) and (\ref{eq:ctp_xi0}) are naturally fulfilled when $\xi=0$ and $\xi=1$. Also, as discussed below, its first derivative must vanish inside of the void $h'(0)=0$. Accordingly, we take the form
\begin{align}
h(\xi) = \xi^2\left(\xi^2-3\xi+3\right) \, .
\label{eq:h_inter}
\end{align}
Recalling Eq. (\ref{eq:concentrations}), we can now define the lithium concentration $c_\Li^\xi$ as
\begin{align}
c_\Li^\xi = h(\xi) c_\Lat^\m\theta_\m \, .\label{eq:lithium_concentration_xi}
\end{align}
Eq. (\ref{eq:lithium_concentration_xi}) shows that, independently of the magnitude of $\theta_\m$, $c_\Li^\xi$ vanishes in the void.\\

Considering these $\xi$-dependent variables, the Helmholtz free energy (\ref{eq:Helmholtz_li_metal}) can be rewritten as
\begin{align}
\Psi^\bulk = \int_V \left\{ c_\Lat^\xi RT\left[\theta_\m \ln{\frac{\theta_\m}{\theta_\m^0}}+ \left(1-\theta_\m\right)\ln{\frac{1-\theta_\m}{1-\theta_\m^0}}\right] + \psi_e^\xi \right\} \,\mathrm{d}V \, .
\end{align}
We further introduce an interfacial energy of the following form
\begin{align}
\Psi^\inter = \int_V \left( wg(\xi) + \frac{1}{2}\kappa|\bm\nabla\xi|^2 \right)\mathrm{d}V \, , \label{eq:psi_interface}
\end{align}

\noindent where $\kappa$ is the gradient energy coefficient and the first term on the right hand side of (\ref{eq:psi_interface}) is a double-well function with two minima at $\xi=0$ and $\xi=1$,
\begin{align}
g(\xi) = \xi^2(1-\xi)^2 \, ,
\end{align}

\noindent where $w/16$ is the barrier height. The total free energy is then defined as the sum of the bulk and interfacial terms: $\Psi = \Psi^\bulk + \Psi^\inter$.

\subsection{Kinematic and constitutive equations}

The proposed deformation-diffusion-phase field model is described by three primal kinematic variables: the displacement field vector $\mathbf{u} ( \mathbf{x}, t)$, the lithium occupancy $\theta_\m (\mathbf{x}, t)$, and the phase field order parameter $\xi ( \mathbf{x}, t)$. These variables respectively characterise the mechanical response, lithium diffusion (vacancy and adatom diffusion), and the evolution of the void-anode interface, as dictated by site nucleation and annihilation. In this subsection, we provide a description of the related kinematic and constitutive equations.

\subsubsection{Mechanical relations}
\label{sec:MechRelations}

\noindent The deformation of the lithium metal electrode is characterised by a total strain rate $\dot{\bm{\varepsilon}}$, defined as
\begin{align}
 \dot{\bm\varepsilon} = \frac{1}{2}\left(\bm\nabla \dot{\mathbf{u}}+\bm\nabla \dot{\mathbf{u}}^\mathrm{T}\right) \, ,
 \end{align} 
\noindent where the superposed dot indicates differentiation with respect to time. The total strain rate can be additively decomposed into its elastic, viscoplastic and chemical components as follows
\begin{align}
\dot{\bm\varepsilon}= \dot{\bm{\varepsilon}}_e + \dot{\bm{\varepsilon}}_v + \dot{\bm{\varepsilon}}_c \, .
\end{align}

The chemical strain rate is the strain variation induced by the transport of Li atoms within the electrode as, due to the different partial molar volumes of occupied and vacant lattice sites, lattice distortion takes place when Li atoms diffuse to occupy a vacancy. We define the chemical strain rate tensor as
\begin{align}
\dot{\bm{\varepsilon}}_c = \frac{1}{3}\frac{\partial (c_\Li^\xi-c_\mathrm{ref}^\xi)}{\partial t}\left(\Omega_\Li-\Omega_\vac\right)\bm 1 \, ,
\end{align}

\noindent where $\bm{1}$ is the second-order unit tensor and $c_\mathrm{ref}^\xi$ is the stress-free concentration, which equals $c_\mathrm{ref}^\xi = h (\xi) c_\Lat^\m\theta_\m^0$. The term $(\Omega_\Li-\Omega_\vac)$ results from the consideration of the substitutional nature of Li diffusion. Recalling Eq. (\ref{eq:lithium_concentration_xi}), we apply the chain rule to the concentration rate, which yields
\begin{align}
\frac{\partial (c_\Li^\xi-c_\mathrm{ref}^\xi)}{\partial t} = \dot hc_\Lat^\m(\theta_\m-\theta_\m^0) + hc_\Lat^\m\dot\theta_\m \, . \label{eq:rate_concentration}
\end{align}

Inspection of (\ref{eq:rate_concentration}) reveals that the rate of concentration has two contributions: site nucleation/annihilation ($\dot h$) and lithium diffusion in the lattice ($\dot{\theta}_\m$). Here, one should note that Li insertion or extraction is only allowed to take place at lattice sites (regions with $\xi>0$) and accordingly the chemical strain rate should vanish in the void. This is automatically satisfied in Eq. (\ref{eq:rate_concentration}) because $h(\xi)$ is defined such that $h(0) = h^\prime(0) = 0$, see Eq. (\ref{eq:h_inter}).\\ 
 
The viscoplastic strain rate $\dot{\bm{\varepsilon}}_v$ is defined following Anand's model \citep{Narayan2018,Anand2019a}, such that
\begin{align}
\dot{\bm{\varepsilon}}_v =F_\mathrm{cr}\frac{3}{2}\frac{\bm\sigma_\dev^\xi}{\sigma_e^\xi},\label{eq:evolotion_viscous}
\end{align}

\noindent where $\bm\sigma_\dev^\xi$ and $\sigma_e^\xi$ denote the deviatoric stress tensor and the von Mises effective stress, respectively. They are defined as
\begin{align}
&\bm\sigma_\dev^\xi = \bm\sigma^\xi - \frac{1}{3}\left(\mathrm{tr}\bm\sigma^\xi\right)\bm 1,\label{eq:deviatoric_stress_def}\\
&\sigma_e^\xi = \sqrt{\frac{3}{2}\bm\sigma_\dev^\xi:\bm\sigma_\dev^\xi}\label{eq:von_Mises_stress_def}
\end{align}
where $\mathrm{tr}\bm\sigma^\xi$ is the trace of the stress tensor $\bm\sigma^\xi$.
The equivalent plastic shear strain-rate $F_\mathrm{cr}$ is given by
\begin{align}
F_\mathrm{cr} = A \exp \left( -\frac{Q}{RT} \right) \left[\sinh\left(\frac{\sigma_e^\xi}{S_a}\right)\right]^{\frac{1}{m}}
\end{align}
where $A$ is a pre-exponential factor, $Q$ is the activation energy, and $m$ is the strain rate sensitivity exponent ($0<m\leq1$). The flow resistance rate $\dot{S}_a$ is defined as follows,
\begin{align}
& \dot{S}_a = H_0\left|1-\frac{S_a}{S_a^\ast}\right|^a\mathrm{sign}\left(1-\frac{S_a}{S_a^\ast}\right)F_\mathrm{cr}, \,\,\,\,\,\,\,\,\, \text{with} \,\,\,\,\,\,\,\,\, S_a^\ast = S_0\left(\frac{F_\mathrm{cr}}{A \exp \left( -\frac{Q}{RT}\right)} \right)^n \, .
\end{align}

\noindent Here, $H_0$, $a$, $n$ and $S_0$ are strain hardening parameters (see \citealp{Anand2019a}). 

Regarding the elastic part, the elastic constants are degraded by the phase field to ensure that a zero stiffness response is attained in the void. Accordingly, we interpolate the elastic moduli as
\begin{align}
G^\xi = h (\xi) G \, , \quad\text{and} \quad K^\xi = h(\xi) K
\end{align}
where $G$ and $K$ are the shear and bulk modulus of the lithium metal, respectively. Along the same lines, we construct the elastic strain energy as 
\begin{align}
\psi_e^\xi = G^\xi {\bm{\varepsilon}}_e : {\bm{\varepsilon}}_e + \frac{1}{2}\left(K^\xi-\frac{2}{3}G^\xi\right)\left(\mathrm{tr}{\bm{\varepsilon}}_e \right)^2 \, ,
\end{align}
\noindent and the definition of the stress tensor $\bm\sigma^\xi$ readily follows as,
\begin{align}
\bm\sigma^\xi = 2G^\xi {\bm{\varepsilon}}_e + \left(K^\xi-\frac{2}{3}G^\xi\right)\left(\mathrm{tr} {\bm{\varepsilon}}_e \right){\mathbf 1}.\label{eq:stress_expression}
\end{align}

Combining Eqs. (\ref{eq:deviatoric_stress_def}), (\ref{eq:von_Mises_stress_def}) and (\ref{eq:stress_expression}), we can define the deviatoric and effective von Mises stresses, as required to calculate the viscoplastic strain, giving
\begin{align}
&\bm\sigma_\dev^\xi = 2G^\xi {\bm{\varepsilon}}_e -\frac{2}{3}G^\xi\left(\mathrm{tr} {\bm{\varepsilon}}_e \right){\mathbf 1},\\
&\sigma_e^\xi = \sqrt{\frac{3}{2}\bm\sigma_\dev^\xi:\bm\sigma_\dev^\xi} = G^\xi \sqrt{6 {\bm{\varepsilon}}_e :{\bm{\varepsilon}}_e + 4\left(\mathrm{tr}\bm\varepsilon_e\right)^2} \, .
\end{align}

\subsubsection{Chemical potential for lithium diffusion}

We proceed to define the chemical potential for the diffusion of lithium atoms within the anode. The evolution of lithium per unit area is described by the time derivative of lithium concentration $c_\Li^\xi$, as defined in (\ref{eq:lithium_concentration_xi}). The work conjugate of $c_\Li^\xi$ is the chemical potential of lithium atoms, which can be derived as
\begin{align}
\mu_\Li^\xi =\frac{1}{c_\Lat^\xi}\frac{\delta\Psi}{\delta\theta_\m} =  - RT\ln{\frac{\theta_\m^0}{1-\theta_\m^0}} + RT\ln{\frac{\theta_\m}{1-\theta_\m}} - \left(\Omega_\Li-\Omega_\vac\right)\sigma_h^\xi\label{eq:mu_Li_xi} \, ,
\end{align}
where $\sigma_h^\xi$ is the hydrostatic stress, given by $\sigma_h^\xi = K^\xi\left(\mathrm{tr}\bm\varepsilon_e\right)$.

\subsubsection{Chemical potential for lithium lattice sites}

The evolution of lattice sites is characterised by the phase field order parameter $\xi$. The work conjugate of $\xi$, which acts as the driving force for the phase field evolution, can be derived as follows,
\begin{align}
\mu_\xi = \frac{\delta\Psi}{\delta\xi} = RT c_\Lat^\m h^\prime\ln{\frac{1-\theta_\m}{1-\theta_\m^0}} + \Omega_\vac c_\Lat^\m h^\prime\psi_e + wg^\prime - \kappa\nabla^2\xi.\label{eq:mu_xi}
\end{align}

It can be seen that, if the interfacial and pressure terms are ignored, $\mu_\xi$ differs from the lattice site chemical potential (\ref{eq:mu_L}) only by the factor $h^\prime c_\Lat^\m$. 

\subsection{Governing equations}

\begin{figure}[H]
\centering
\includegraphics[width=0.8\linewidth]{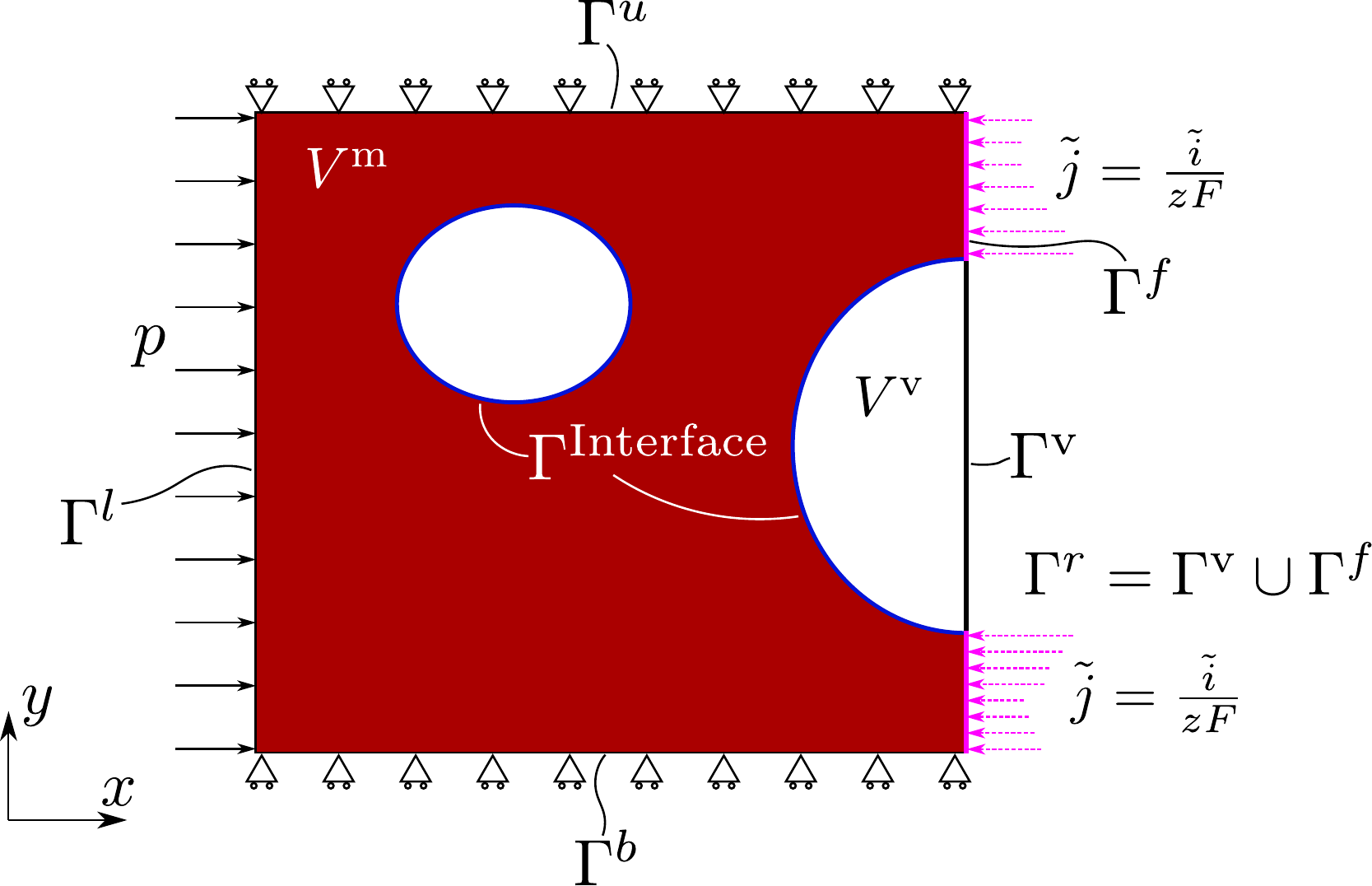}
\caption{Illustration of different domains, boundaries and interfaces of the problem.}\label{fig:problem_setup}
\end{figure}

We shall now derive the governing equations. As shown in Fig. \ref{fig:problem_setup}, the lithium electrode domain is enclosed by four boundaries $\Gamma^l$, $\Gamma^r$, $\Gamma^u$ and $\Gamma^b$ on the left, right, upper and lower edges, respectively. The current collector and the solid electrolyte lie on the left and right sides of the domain of interest, respectively. The upper and lower edges are electronically isolated, chemically impermeable and mechanically confined. The electrode area is further subdivided into two parts: the lithium metal $V^\m$ and the void $V^\vac$. We then identify two more interfaces: the one between the lithium metal and the void $\Gamma^\inter$, and that between the void and the solid electrolyte $\Gamma^\vac$. Using the phase field paradigm, $\Gamma^\inter$ has been replaced by a diffuse interface, characterised by the phase field order parameter $\xi$ and its gradient. Following experimental observations, we assume that voids can nucleate from the electrolyte side. Under this assumption, the solid electrolyte can be in contact with both the void and the lithium metal. We denote the interface with the void as $\Gamma^\vac$ and the interface with lithium metal as $\Gamma^f$. Then, $\Gamma^\vac\cup\Gamma^f=\Gamma^r$. We choose lithium occupancy $\theta_\m$, displacements $\mathbf{u}$ and phase field order parameter $\xi$ as independent field variables, which are defined in the complete domain $V$.

\subsubsection{Field equations}\label{subsubsec:field_equations}

In order to capture the transport of Li atoms, the annihilation and creation of lithium lattice sites and the deformation of the Li metal, we derive the following three sets of governing equations. \\

\noindent \textbf{Li transport}. The evolution of the Li concentration is governed by the following mass transfer equation
  \begin{align}
\frac{\partial c_\Li^\xi}{\partial t} = -\bm\nabla\cdot \mathbf{j}_\Li^\xi \, , \label{eq:evolution_concentration_brief}
\end{align}
\noindent where $\mathbf{j}_\Li^\xi$ is the lithium flux, which is defined as
\begin{align}
\mathbf{j}_\Li^\xi = - \frac{D_\mathrm{eff}c_\Li^\xi}{RT}\bm\nabla\mu_\Li^\xi \, ,\label{eq:flux_Li_define}
\end{align}

\noindent with $D_\mathrm{eff}$ being the effective diffusion coefficient. Combining \cref{eq:mu_Li_xi,eq:flux_Li_define} yields
\begin{align}
\mathbf{j}_\Li^\xi = -\frac{D_\mathrm{eff}h (\xi) c_\Lat^\m}{1-\theta_\m}\bm\nabla\theta_\m + \frac{D_\mathrm{eff}h(\xi) c_\Lat^\m\theta_\m(\Omega_\Li-\Omega_\vac)}{RT}\bm\nabla\sigma_h^\xi\ . \label{eq:flux_xi}
\end{align}

\noindent Hence, the lithium flux is driven by the gradients of lithium occupancy and hydrostatic stress. The flux must be zero inside of the void and this is naturally captured in the formulation, as $h(\xi=0)=0$. Also, the gradient terms vanish when $\xi=0$, as the lithium occupancy is constant and the void carries no stress. Substituting Eqs. (\ref{eq:lithium_concentration_xi}) and (\ref{eq:flux_xi}) into Eq. (\ref{eq:evolution_concentration_brief}), the governing equation for mass diffusion becomes
\begin{align}
hc_\Lat^\m\frac{\partial \theta_\m}{\partial t} + c_\Lat^\m\theta_\m h^\prime\frac{\partial\xi}{\partial t} = \bm\nabla\cdot\frac{D_\mathrm{eff}hc_\Lat^\m}{1-\theta_\m}\bm\nabla\theta_\m-\bm\nabla\cdot\frac{D_\mathrm{eff}c_\Li^\xi(\Omega_\Li-\Omega_\vac)}{RT}\bm\nabla\sigma_h^\xi.
\end{align}
Dividing by $c_\Lat^\m$ on both sides and rearranging, the chemical balance can be formulated as,
\begin{align}
h\frac{\partial\theta_\m}{\partial t} + \theta_\m h^\prime\frac{\partial\xi}{\partial t} = \bm\nabla\cdot\frac{D_\mathrm{eff}h}{1-\theta_\m}\bm\nabla\theta_\m-\bm\nabla\cdot\frac{D_\mathrm{eff}h\theta_\m(\Omega_\Li-\Omega_\vac)}{RT}\bm\nabla\sigma_h^\xi\label{eq:govern_diffusion}
\end{align}

Inspection of Eq. (\ref{eq:govern_diffusion}) reveals that surface diffusion plays a dominant role, as $\sigma_h^\xi$ varies significantly along the interface. This is in agreement with the terrace-ledge-kink model and the lower activation barriers reported for self-diffusion along surfaces \citep{Krauskopf2019}. It is also worth emphasising the different nature of stress-assisted Li diffusion and creep, with the former being driven by the hydrostatic stress gradient (\ref{eq:govern_diffusion}) while the latter depends on the deviatoric stress (\ref{eq:evolotion_viscous}).\\

\noindent \textbf{Evolution of lattice sites}. The nucleation and growth of voids are driven by the annihilation and creation of lattice sites; here described by a novel phase field formulation. The density of lattice sites is not a conserved quantity and its evolution is driven by the free energy of the system. Accordingly, the evolution for the phase field parameter $\xi$ follows an Allen--Cahn-type equation
\begin{align}
\frac{\partial\xi}{\partial t} = -L\mu_\xi = -\frac{LRT h^\prime}{\Omega_\Lat}\ln{\frac{1-\theta_\m}{1-\theta_\m^0}} - \frac{L\Omega_\vac h^\prime}{\Omega_\Lat}\psi_e - Lwg^\prime + L\kappa\nabla^2\xi \, , \label{eq:govern_evolution_xi}
\end{align}

\noindent where $L$ is the so-called phase field mobility or kinetic parameter.\\

\noindent \textbf{Mechanical deformation}. In the absence of body forces and neglecting the role of inertia, the mechanical behaviour of the electrode is characterised by the balance of linear momentum:
\begin{align}
\bm\nabla\cdot\bm\sigma^\xi =\bm 0 \, , \label{eq:govern_force}
\end{align}

\noindent where the $\xi$-dependent stress tensor is used to capture the loss of stiffness associated with voided regions ($\xi=0$). 

\subsubsection{Boundary conditions}

We proceed to discuss the boundary conditions for the couple deformation-diffusion-phase field problem. For the sake of clarity, these are divided in three sets, as in the description of the local force balances.\\ 

\noindent \textbf{Li transport}. As shown in Fig. \ref{fig:problem_setup}, from the chemical viewpoint there is only one relevant boundary; $\Gamma^r$, the one in contact with the electrolyte, through which lithium ions can penetrate \textit{via} the following electrochemical reaction
\begin{align}
\ce{Li+} + \ce{e-}\rightleftharpoons \ce{Li}. \label{eq:chemical_reaction_Li_disslution}
\end{align}
All other three boundaries are impermeable to lithium atoms. The arising current density $\tilde i$ can be obtained from a Butler--Volmer equation based on the above electrochemical reaction. It can also be prescribed as a distributed current density. As sketched in Fig. \ref{fig:problem_setup}, no reactants are present in the void and consequently the current must vanish along the void-electrolyte interface. In the remaining parts of the electrode-electrolyte interface the flux should be proportional to the current density, such that the boundary condition is given by
\begin{align}
&-\mathbf{j}_\Li^\xi\cdot \mathbf{n} = \frac{\tilde i}{zF}&&\text{on}\quad\Gamma^r \label{eq:flux1}\\
&-\mathbf{j}_\Li^\xi\cdot \mathbf{n} = 0 &&\text{on}\quad\Gamma^l\cup\Gamma^u\cup\Gamma^b.
\end{align}
where $F$ is the Faraday constant and $z$ is the charge number of the ion, equal to 1 for Li. One should note that if Eq. (\ref{eq:govern_diffusion}) is used as the local force balance, both sides in (\ref{eq:flux1}) should be divided by $c_\Lat^\m$. Since $c_\Lat^\m\Omega_\Lat=1$, the boundary condition can be reformulated as
\begin{align}
&-\frac{\mathbf{j}_\Li^\xi}{c_\Lat^\m}\cdot \mathbf{n} = \frac{\tilde i\Omega_\Lat}{zF}&&\text{on}\quad\Gamma^r\\\label{eq:flux_boundary_applied}
&-\mathbf{j}_\Li^\xi\cdot \mathbf{n} = 0 &&\text{on}\quad\Gamma^l\cup\Gamma^u\cup\Gamma^b.
\end{align}

We emphasise that the model comprises two classes of reactions: (i) Li dissolution/deposition, shown in Eq. (\ref{eq:chemical_reaction_Li_disslution}) and captured through the interfacial flux and current density, see Eq. (\ref{eq:flux1}); and (ii) vacancy annihilation/nucleation, which is modelled through Eq. (\ref{eq:govern_evolution_xi}). Reaction (i) can only occur where the Li metal anode meets the solid electrolyte, where a vacancy can be generated (stripping) or consumed (plating) as a result, see Fig. \ref{fig:SketchPhenomena}. Reaction (ii) can occur anywhere inside the Li metal anode as long as the chemical potential allows for it. It may also occur along with Reaction (i). For instance, for a perfect Li-electrolyte interface without voids, Li atoms at the interface are oxidised and insert into the electrolyte uniformly along the interface during stripping, leaving a layer of vacancies that may annihilate into a layer of voids; as a consequence, the lithium anode is peeled off from the electrolyte completely if no stack pressure is applied to maintain the contact.\\

\noindent \textbf{Evolution of lattice sites}. The boundary condition for the phase field order parameter $\xi$ is a natural boundary condition for all sides, viz.
\begin{align}
\bm\nabla\xi\cdot \mathbf{n} = 0  &&\text{on}\quad\Gamma^l\cup\Gamma^u\cup\Gamma^b\cup\Gamma^r.
\end{align}

\noindent \textbf{Mechanical deformation}. Unlike the case of the diffusion problem, the only relevant boundary from a mechanical point of view is the left side of the electrode, $\Gamma^l$. We assume all boundaries are traction free in the tangential direction. In the normal direction to the boundary, pressure can be applied to minimise voiding and maximise the contact between the electrode and the electrolyte. Accordingly, the boundary conditions in the normal direction are defined as 
\begin{align}
&\mathbf{n}\cdot \bm\sigma^\xi\cdot \mathbf{n}= -p_\mathrm{applied}\,\,\,\,\text{ or }\,\,\,\, \mathbf{u}\cdot\mathbf{n} = \dot{u}_\mathrm{applied}t\quad&&\text{on}\quad\Gamma^l,\\
& \mathbf{u}\cdot\mathbf{n} = 0 &&\text{on}\quad\Gamma^u\cup\Gamma^b\cup\Gamma^r.
\end{align}

\subsubsection{Initial conditions}

Finally, the initial conditions are defined as,
\begin{align}
&\xi = 1, \quad \theta_\m = \theta_m^0 && \text{in} \quad V^\m\\
&\xi = 0, \quad \theta_\m = \theta_m^0 && \text{in} \quad V^\vac.
\end{align}

In all computations, the phase field distribution is allowed to equilibrate before applying any loading.

\subsection{Coupling with the electro-mechanical behaviour of the solid electrolyte}
\label{subsec:model_electrolyte}

\begin{figure}[H]
\centering
\includegraphics[width=0.8\linewidth]{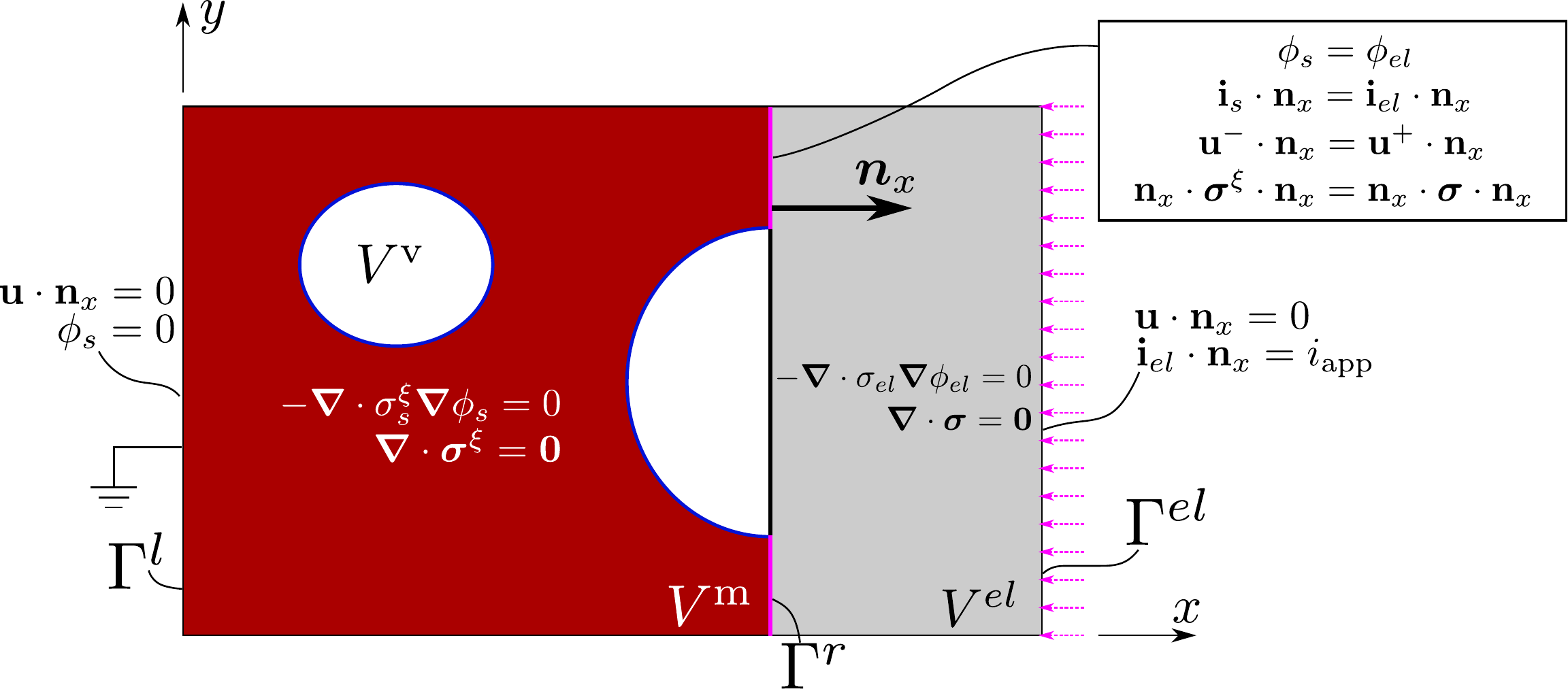}
\caption{Complete electro-chemo-mechanical model, coupling the deformation-diffusion-phase field behaviour of the Li metal electrode with the electro-mechanical behaviour of the solid electrolyte.}\label{fig:problem_setup_electrolyte}
\end{figure}

To enable the prediction of current hot spots, which act as dendrite nucleation sites, we extend our electrode deformation-diffusion-phase field model to capture the electro-mechanical behaviour of the electrolyte. As shown in Fig. \ref{fig:problem_setup_electrolyte}, the electrolyte occupies a volume $V^{el}$ and is in contact with the right hand side of the electrode at the interface $\Gamma^r$. The additional governing equations and boundary conditions for the coupled problem are summarised here and in Fig. \ref{fig:problem_setup_electrolyte}; the reader is referred to \citet{Zhao2022} for a more comprehensive description of solid electrolyte behaviour. 

The following assumptions underpin our model: (i) the electrolyte is a single ion conductor, where Li ion is the only charge carrier, (ii) the concentration of Li ion in the electrolyte remains constant in time, and (iii) the diffusion of Li ions within the interstitials of the electrolyte will not distort the electrolyte lattice. These assumptions exempt us from solving the lithium concentration field explicitly and allow us to focus on the electric and mechanical fields in the electrolyte. The current delivered by \ce{Li+} is dictated by Ohm's law and reads, 
\begin{align}
\mathbf{i}_{el}=-\sigma_{el} \bm{\nabla} \phi_{el} \, ,\label{eq:electrolyte_current}
\end{align}

\noindent where $\mathbf{i}_{el}$ is the electric current and $\sigma_{el}$ is the electric conductivity. Upon further assuming that no charge is present, one reaches
\begin{align}
\bm \nabla \cdot \mathbf{i}_{el} = 0.\label{eq:electrolyte_poisson}
\end{align}

Mechanically, we assume that the electrolyte behaves in a linear elastic manner. Thus, the displacement $\mathbf{u}$ in the electrolyte is described by the Navier-Lam\`e equation
\begin{align}
\left(\lambda+G\right)\bm\nabla\left(\bm\nabla\cdot\mathbf{u}\right) + G\bm\nabla^2\mathbf{u} = \mathbf{0} \, ,
\end{align}
where $\lambda$ and $G$ are the Lam\`e constants of the electrolyte.
%


We shall now turn our attention to the governing equations in the electrode. The chemical and mechanical governing equations have been introduced in Section \ref{subsubsec:field_equations}. Now we derive the governing equation for the electric field. The current in the Li metal electrode anode is conducted by electrons. Since the conductivity of the Li metal ($\sigma_s$) is several orders of magnitude greater than that of the electrolyte, we can adopt the common assumption of a constant electric potential within the electrode. However, the air in the void can act as an electron insulator with zero conductivity, influencing the distribution of electric potential in the electrode. Therefore, we define a phase field dependent conductivity 
\begin{align}\label{eq:PFconduct}
\sigma_s^\xi = f(\xi)\sigma_s \, .
\end{align}

The choice of $f(\xi)$ must satisfy the following requirements: $f(0)=0$, $f(1)=1$, and $f'(0)=0$. Here, we choose to adopt the following higher-order form to achieve a steep conductivity change near the interface,
\begin{equation}
    f(\xi)=\xi^{15}(\xi^4-3\xi^2+3) \, .
\end{equation}
The current density ($\mathbf{i}_s$) and electric potential ($\phi_s$) in the electrode then read
\begin{align}
&\mathbf{i}_s=-\sigma_s^\xi \bm{\nabla} \phi_s,\\
&\bm \nabla \cdot \mathbf{i}_s = 0. \label{eq:OhmLaw}
\end{align}

Finally, the boundary and interface conditions are defined. As shown in Fig. \ref{fig:problem_setup_electrolyte}, two boundaries are identified: the left side of the electrode domain ($\Gamma^l$) and the right side of the electrolyte ($\Gamma^{el}$). The electrode is in contact with the current collector on the left, which ensures that the electrode potential vanishes. The electrolyte is in contact with a counter electrode on the right, which supplies a uniformly distributed current density $i_\mathrm{app}$. The interface impedance is disregarded at the electrode-electrolyte interface $\Gamma^r$ and thus both current density and electric potential are continuous. Mechanically, we fix all normal displacements on $\Gamma^l$ and $\Gamma^{el}$. At the interface $\Gamma^r$, continuity conditions apply. Accordingly, the boundary and interface conditions read

%
\begin{align}
& \phi_s = 0, 
&& \mathbf{u}\cdot\mathbf{n}_x = 0, &&\text{on}\quad\Gamma^l,\label{eq:electrolyte_boundary_1}\\
& \phi_s = \phi_{el},\,
&& \mathbf{u}^-\cdot\mathbf{n}_x = \mathbf{u}^+\cdot\mathbf{n}_x, \, \nonumber \\
& \mathbf{i}_s\cdot \mathbf{n}_x = \mathbf{i}_{el}\cdot \mathbf{n}_x , \, 
 && 
 \mathbf{n}_x\cdot\bm\sigma^\xi\cdot\mathbf{n}_x = \mathbf{n}_x\cdot\bm\sigma\cdot\mathbf{n}_x, &&
 \text{on}\quad\Gamma^r,\label{eq:electrolyte_boundary_2}\\
& \mathbf{i}_{el}\cdot \mathbf{n}_x = i_\mathrm{app}, 
&& \mathbf{u}\cdot\mathbf{n}_x = 0, &&\text{on}\quad\Gamma^{el}\label{eq:electrolyte_boundary_3}
\end{align}
where $\mathbf{n}_x$ is the normal vector along the $x$ direction. Superscripts $^-$ and $^+$ denote the left and right side of the interface $\Gamma^r$, respectively.\\

\noindent \textbf{Remark.} In our formulation, we encapsulate the four kinetic events that govern void evolution: (i) the rate of Li dissolution/deposition, (ii) Li diffusion, (iii) the nucleation/annihilation of vacancies, and (iv) creep. We independently incorporate these phenomena through boundary/interface conditions (for (i)), field governing equations (for (ii) and (iii)) and constitutive relations (for (iv)). As described in \ref{subsec:interface_velocity}, in the absence of applied current and bulk diffusion, we can derive a simplified model and estimate the interface velocity. The resulting expression for the interface velocity highlights the importance of the lattice annihilation-induced pressure source term inside the interface, and the similarities with traditional models of surface diffusion \citep{Chuang1979,NEEDLEMAN1983}.

\section{Numerical experiments}
\label{Sec:Results}

The theoretical framework described in Section \ref{Sec:Theory} is numerically implemented using the finite element method. Specifically, the commercial finite element package COMSOL Multiphysics is used. Plane strain conditions are assumed, time integration is carried out using a backward Euler method, and quadratic quadrilateral elements are used for discretising the electrode and electrolyte domains. A mesh sensitivity analysis is conducted in all the computations, with the characteristic element length in the regions of void evolution being at least ten times smaller than the inteface thickness $\ell$, which is sufficient to ensure mesh-independent results \citep{PTRSA2021}. As derived in \ref{App:interface_thickness}, the interface length equals,
\begin{equation}
    \ell = \sqrt{\frac{8\kappa}{w}} \, .
\end{equation}

The number of degrees-of-freedom employed in the various boundary value problems examined ranges from 25 to 30 million. To achieve mesh-objective results, the finite element mesh has to be sufficiently refined along the electrode-electrolyte and Li metal-void interfaces to resolve the gradients of the phase field order parameter and the local current.\\

We begin, in Section \ref{Sec:CalibrationLi}, by describing our choice of electrode-electrolyte system, presenting the phase field and material parameters, and calibrating the viscoplastic constitutive behaviour of metallic lithium with the uniaxial tension tests conducted at various strain rates by \citet{LePage2019}. Then, numerical experiments are conducted on a single void model (Section \ref{Sec:SingleVoid}) to investigate: (i) the role of plating and stripping in driving void evolution (Section \ref{Sec:PlattingStripping1Void}), (ii) the sensitivity to the applied current density and the phase field mobility parameter (Section \ref{Sec:LANDCurrentDensity}), and (iii) the interplay between creep and vacancy diffusion (Section \ref{Sec:CreepVSdiffusion}). Subsequently, a full-scale model is developed, which includes multiple voids and where several charge/discharge cycles are simulated to mimic realistic conditions (Section \ref{Sec:FullModel}). This realistic model is used to gain insight into the important role of the applied pressure and the results obtained are discussed in the context of experimental observations. In these boundary value problems, voids are introduced to mimic the non-ideal solid-solid contact between the Li anode and the ceramic electrolyte. Gaps along the anode-electrolyte interface can arise from multiple sources, including the pores that nucleate at impurities, volume change differences between charged and discharged electrodes \citep{Devaux2015}, and the defects inherent to the manufacture of ceramic materials \citep{Porz2017}. In the absence of initial defects (a perfect, void-free interface), the model predicts uniform Li dissolution or deposition for stripping and plating, respectively. For the case of stripping, this leads to the formation of a layer where $\xi=0$, due to the role of the stripping flux \textit{via} Eq. (\ref{eq:flux1}) and vacancy annihilation. This layer soon stabilises at a thickness that scales with $\ell$, capturing the arrest of the reaction (\ref{eq:chemical_reaction_Li_disslution}) as a gap opens between the Li metal anode and the electrolyte.

\subsection{Phase field and material parameters for a Li anode - LLZO electrolyte system.}
\label{Sec:CalibrationLi}

We choose to conduct our numerical experiments in a cell composed of a metallic lithium anode and a Li$_7$La$_3$Zr$_2$O$_{12}$ (LLZO) solid electrolyte, arguably the most relevant electrode-electrolyte system. The garnet LLZO is a common choice in the experimental and theoretical literature, as it can be synthesized at high relative densities ($>$95\%), exhibits high conductivity and stiffness, and is stable against metallic lithium \citep{Wang2019a,Manalastas2019}. The phase field and material parameters used are shown in Table \ref{tb:elastic parameters}. 

\begin{table}[H]\small
\centering
\caption{Phase field and material parameters for a Li anode - LLZO electrolyte system.}
\label{tb:elastic parameters}
\setlength{\tabcolsep}{10mm}{
\begin{tabular}{ll}
\hline
Parameter   & Magnitude   \\ \hline
Effective diffusion coefficient, $D_{\mathrm{eff}}$ [\si{\meter\squared\per\second}]  & \num{7.5e-13}\ (i) \\
Young's modulus of lithium metal, $E_{\mathrm{Li}}$ [\si{\giga\pascal}]    & \num{4.9}\ (ii)     \\
Young's modulus of LLZO, $E_{\mathrm{LLZO}}$ [\si{\giga\pascal}]    & \num{150}\ (iii)     \\
Poisson's ratio of lithium metal, $\nu_{\mathrm{Li}}$ [-]        & \num{0.38}\ (ii) \\
Poisson's ratio of LLZO, $\nu_{\mathrm{LLZO}}$ [-]        & \num{0.257}\ (iii) \\
Interface kinetics coefficient, $L$ [\si{\meter\squared\per\newton\per\second}]   & \num{1e-9}  \\
Height of the double well potential, $w$ [\si{\newton\per\meter\squared}]     & \num{3.5e6}       \\
Gradient energy coefficient, $\kappa$ [\si{\newton}]  & \num{4.5e-7}        \\
Gas constant, $R$ [\si{\joule\per\mole\per\kelvin}]  & \num{8.314} \\
Absolute temperature, $T$ [\si{\kelvin}]  & \num{298}    \\
Molar volume of lithium, $\Omega_{\mathrm{Li}}$ [\si{\meter\cubed\per\mole}]  & \num{13.1e-6}\ (ii)  \\
Molar volume of vacancies, $\Omega_\vac$ [\si{\meter\cubed\per\mole}]  & \num{6e-6}\ (ii)   \\
Average molar volume of Li lattice sites, $\Omega_\Lat$ [\si{\meter\cubed\per\mole}] & \num{13.1e-6}\ (ii)  \\
Electric conductivity of lithium metal, $\sigma_s$ [\si{\siemens\per\meter}]  & \num{1.1e7}\ (i) \\
Ionic conductivity of solid electrolyte (LLZO), $\sigma_{el}$ [\si{\siemens\per\meter}] & \num{5.5e-6}\ (iv) \\ \hline
\multicolumn{2}{l}{\footnotesize(i) \citet{Chen2015a};\ (ii) \citet{Shishvan2021};\ (iii) \citet{Yu2016b}; \ (iv) \citet{Buschmann2012}}                                           \\ \hline
\end{tabular}}
\end{table}

\begin{figure}[H]
    \centering
    \includegraphics[width=0.8\linewidth]{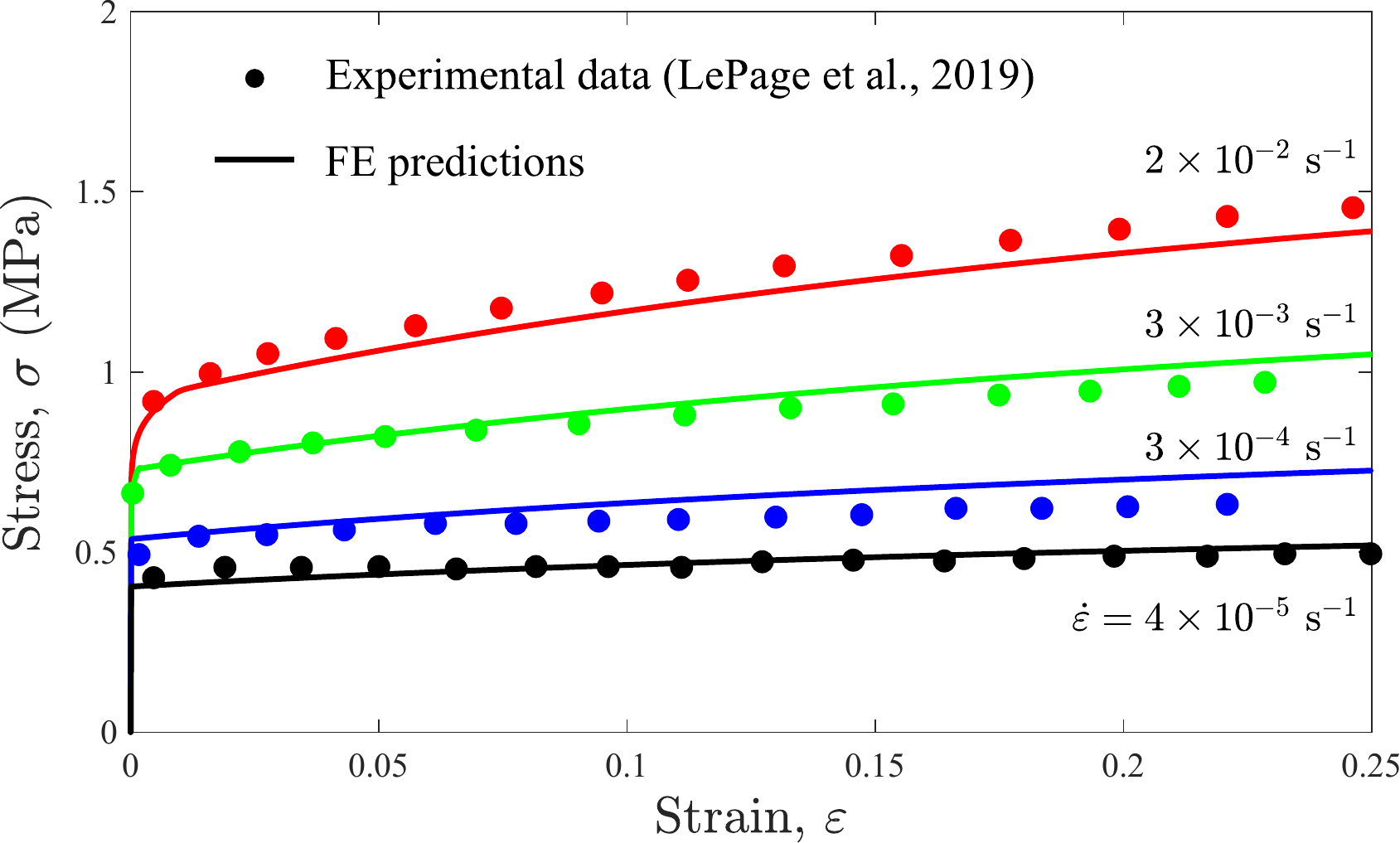}
    \caption{Uniaxial stress $\sigma$ versus strain $\varepsilon$ curves for lithium metal at $T=298$ K and different strain rates ($\dot{\varepsilon}$). Comparison between the experiments by \citet{LePage2019} (symbols) and our numerical predictions (lines) for the parameters listed in Table \ref{tb:visco_parameter}.}
    \label{fig:visco_parameter_cali}
\end{figure}

It remains to define the constitutive parameters of the viscoplastic formulation adopted to characterise the mechanical behaviour of the Li metal anode. As described in Section \ref{sec:MechRelations}, we choose to describe the viscoplastic and creep behaviour of metallic lithium using the model developed by \citet{Anand2019a}. The parameters of the model are calibrated against the experimental work by \citet{LePage2019}. Accordingly, the activation energy is taken to be $Q=37$ kJ/mol and the hardening and strain rate sensitivity coefficients are determined by matching the uniaxial stress-strain curves reported by \citet{LePage2019} at different strain rates and room temperature. The comparison between the experimental data and our numerical results is given in Fig. \ref{fig:visco_parameter_cali}. A good agreement with experiments is obtained for the parameters reported in Table \ref{tb:visco_parameter} and thus these values are adopted in our subsequent calculations.

\begin{table}[H]\small
\centering
\caption{Viscoplastic material parameters for lithium metal.}
\label{tb:visco_parameter}
\setlength{\tabcolsep}{10mm}{
\begin{tabular}{ll}
\hline
Parameter                                                                     & Magnitude            \\ \hline
Pre-exponential factor, $A$ [\si{\per\second}]                         & \num{4.25e4}         \\
Activation energy, $Q$ [\si{\kilo\joule\per\mole}]                            & \num{37}             \\
Strain rate sensitivity exponent, $m$ [-]                                                   & \num{0.15}           \\
Deformation resistance saturation coefficient, $S_0$ [\si{\mega\pascal}]      & \num{2}              \\
Initial value of the flow resistance, $S_a (t=0)$ [\si{\mega\pascal}] & \num{1.1}            \\
Hardening constant, $H_0$ [\si{\mega\pascal}]                                 & \num{10}             \\
Hardening sensitivity, $a$ [-]                                                & \num{2}              \\
Deformation resistance sensitivity, $n$ [-]                                   & \num{0.05}           \\ \hline
\end{tabular}}
\end{table}

\subsection{Single void analysis}
\label{Sec:SingleVoid}

\begin{figure}[H]
    \centering
    \includegraphics[width=0.8\linewidth]{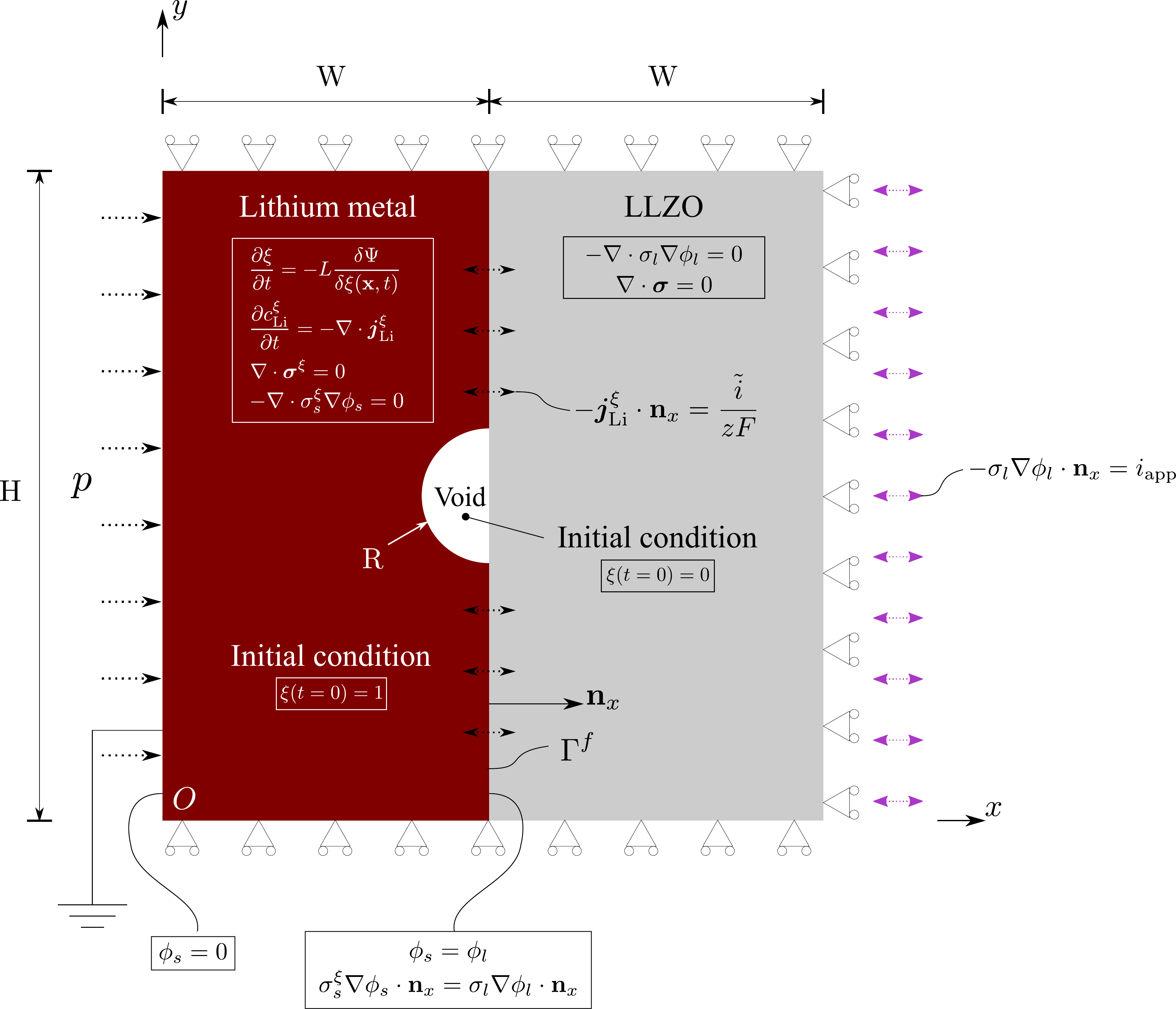}
    \caption{Single void boundary value problem: dimensions, configuration and boundary conditions. The initial conditions and governing equations for the electrolyte and electrode domain are also included.}
    \label{fig:SingleVoidConfig}
\end{figure}

Insight is first gained by simulating the evolution of a single void lying at the interface between the Li metal anode and the electrolyte, see Fig. \ref{fig:SingleVoidConfig}. The electrolyte and the electrode are assumed to have a rectangular shape and equal dimensions, with a height of $\mathrm{H}=\SI{250}{\micro\meter}$ and a width of $\mathrm{W}=0.16\mathrm{H}$, such that the complete domain has dimensions of $250 \times \SI{80}{\micro\meter\squared}$. The void has a perfect semi-circular shape with radius $\mathrm{R}=0.04\mathrm{H}=\SI{10}{\micro\meter}$ initially.

All the edges of the Li metal anode are impermeable to Li flux except for $\Gamma^f$, the electrode-electrolyte interface. Regarding the electric problem, a uniformly distributed current density ($i_\mathrm{app}$) is applied to the right edge of the electrolyte. This current is generally taken to be equal to $|i_\mathrm{app}|=0.1$ \si{\milli\ampere\per\centi\meter\squared} but other magnitudes are also considered to investigate its influence. Mechanically, we constrain the vertical displacements of the bottom and top edges of the electrode and of the electrolyte. The horizontal displacement of the right edge of the electrolyte is also kept equal to zero. This reflects the assumption of composite cathode whose macroscopic volume change is negligible during operation compared with lithium anode. Moreover, the cathode and electrolyte are both assumed much stiffer than lithium metal. As for the left edge, we consider two scenarios: (i) a pressure of $p$ is applied at this boundary to study the influence the applied pressure; or (ii) a fixed displacement is enforced in order to mimic the constraint of a solid shell that wraps the battery cell.

\subsubsection{Void evolution in the electrode under stripping and plating}
\label{Sec:PlattingStripping1Void}

First, we present model predictions of void evolution under stripping and plating conditions. For the sake of clarity, no pressure is applied on the electrode and simulations are presented for one continuous cycle of stripping or plating. Otherwise, the loading conditions correspond to those described above. Currents with the same magnitude (but different sign) are applied for plating and stripping; specifically, $i_\mathrm{app}=0.1$ \si{\milli\ampere\per\centi\meter\squared} during plating and $i_\mathrm{app}=-0.1$ \si{\milli\ampere\per\centi\meter\squared} during stripping. The corresponding \ce{Li+} flux $\bm{j}_\mathrm{Li}^{\xi}$ flowing across the electrode-electrolyte interface is then calculated from the interfacial current density.

\begin{figure}[H]
    \centering
    \includegraphics[width=0.8\linewidth]{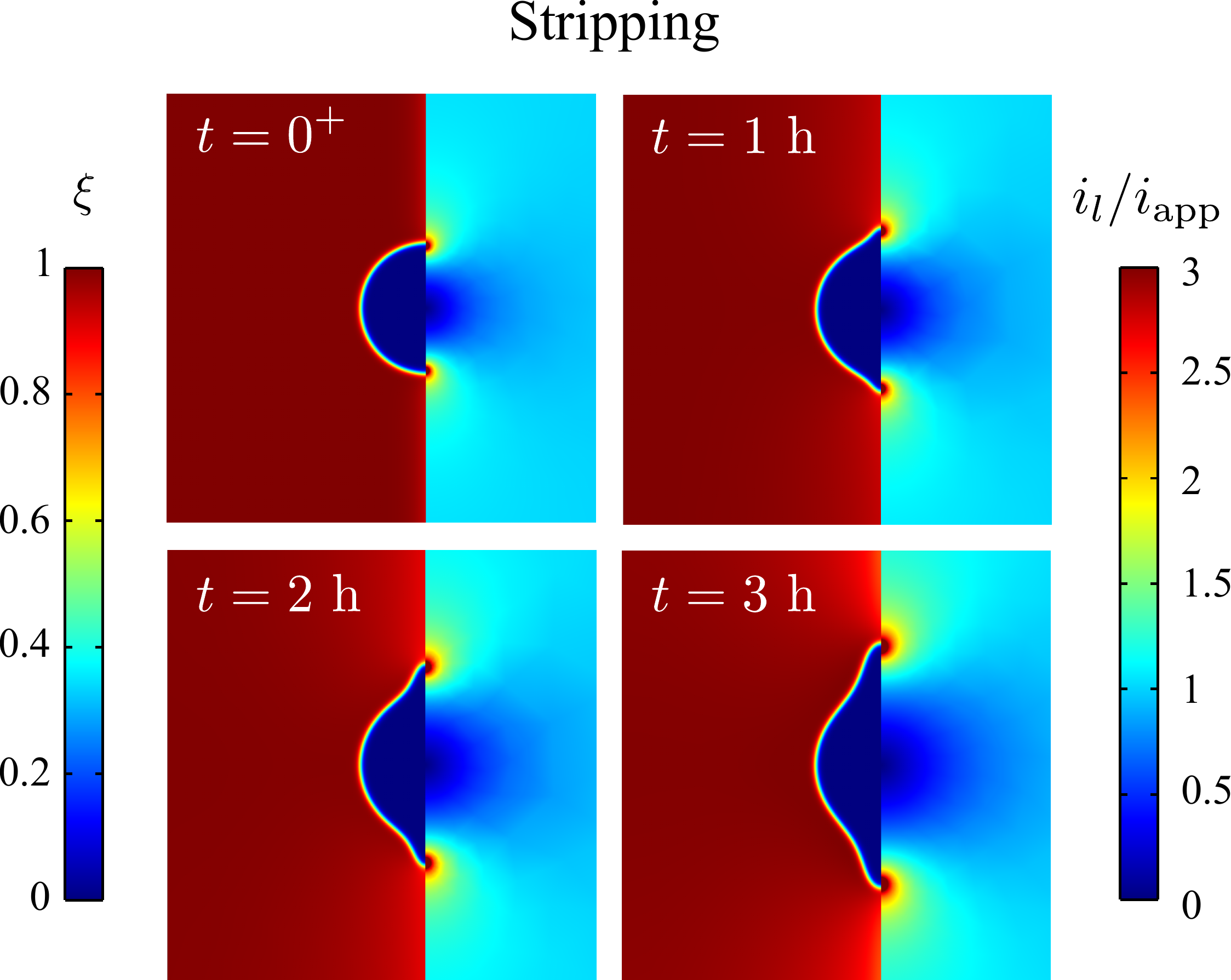}
    \caption{Numerical experiments on a single void model: voiding and local current hot spots under stripping, as characterised by the evolution of the phase field order parameter in the electrode and the current density distribution in the electrolyte.}
    \label{fig:single_void_visco_stripping_contour}
\end{figure}

\begin{figure}[H]
    \centering
    \includegraphics[width=0.8\linewidth]{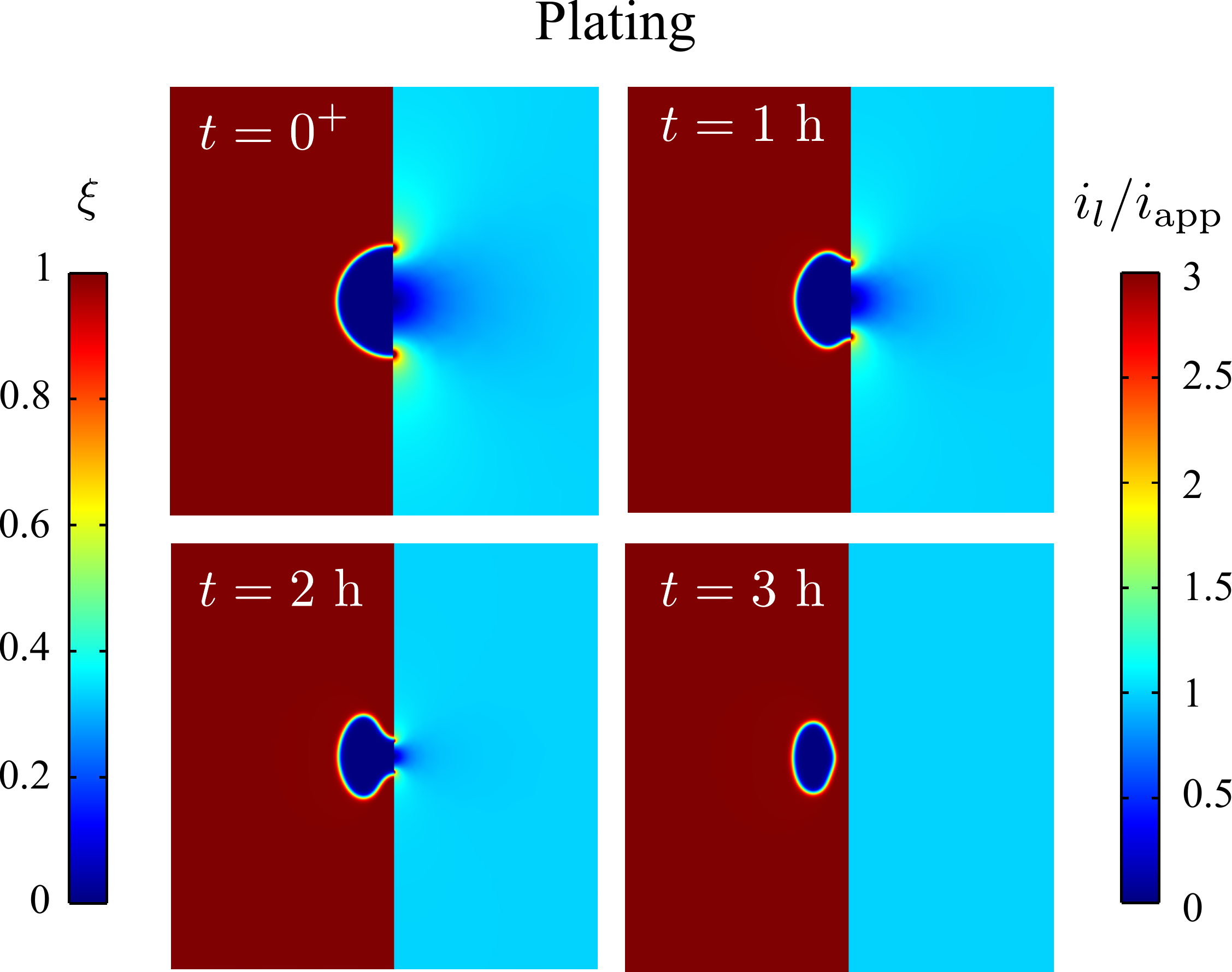}
    \caption{Numerical experiments on a single void model: voiding and local current hot spots under plating, as characterised by the evolution of the phase field order parameter in the electrode and the current density distribution in the electrolyte.}
    \label{fig:single_void_visco_plating_contour}
\end{figure}

The results of the stripping and plating processes are shown in Figs. \ref{fig:single_void_visco_stripping_contour} and \ref{fig:single_void_visco_plating_contour}, respectively. Void evolution is characterised by contours of the phase field order parameter ($\xi$) in the electrode, while the contours in the electrolyte describe the predicted normalised electrolyte current density ($i_l/i_\mathrm{app}$). Consider first the results obtained for the stripping process, Fig. \ref{fig:single_void_visco_stripping_contour}. Our numerical predictions reveal that stripping causes the void to deviate from its original shape, widening along the electrode-electrolyte interface. This is in agreement with experimental observations (see Fig. \ref{fig:introduction}) and occurs due to the faster rate of Li dissolution at the interface, relative to the rate at which Li is supplied from the bulk due to vacancy diffusion. The rate of Li dissolution is further exacerbated by the higher stripping flux in regions of high current density, as per Eq. (\ref{eq:flux_boundary_applied}). These current hot spots arise in the regions where the void, Li metal and solid electrolyte meet, and extend their size as the void grows with increasing stripping time. By accelerating void growth, these regions of high current intensity contribute to reducing the contact area between the electrode and the electrolyte, increasing cell resistance. Moreover, hot spots act as dendrite nucleation sites during the subsequent plating cycle \citep{Manalastas2019}. The emergence of hot spots near the void edges is predicted as a result of the abrupt change in conductivity taking place along the void-Li metal interface, see Eqs. (\ref{eq:PFconduct})-(\ref{eq:OhmLaw}), with the conductivity going from being equal to the Li metal conductivity ($\sigma_s^\xi=\sigma_s$) in the Li phase ($\xi=1$) to zero inside of the void ($\sigma_s^\xi=0$ for $\xi=0$).

\begin{figure}[H]
    \centering
    \includegraphics[width=0.8\linewidth]{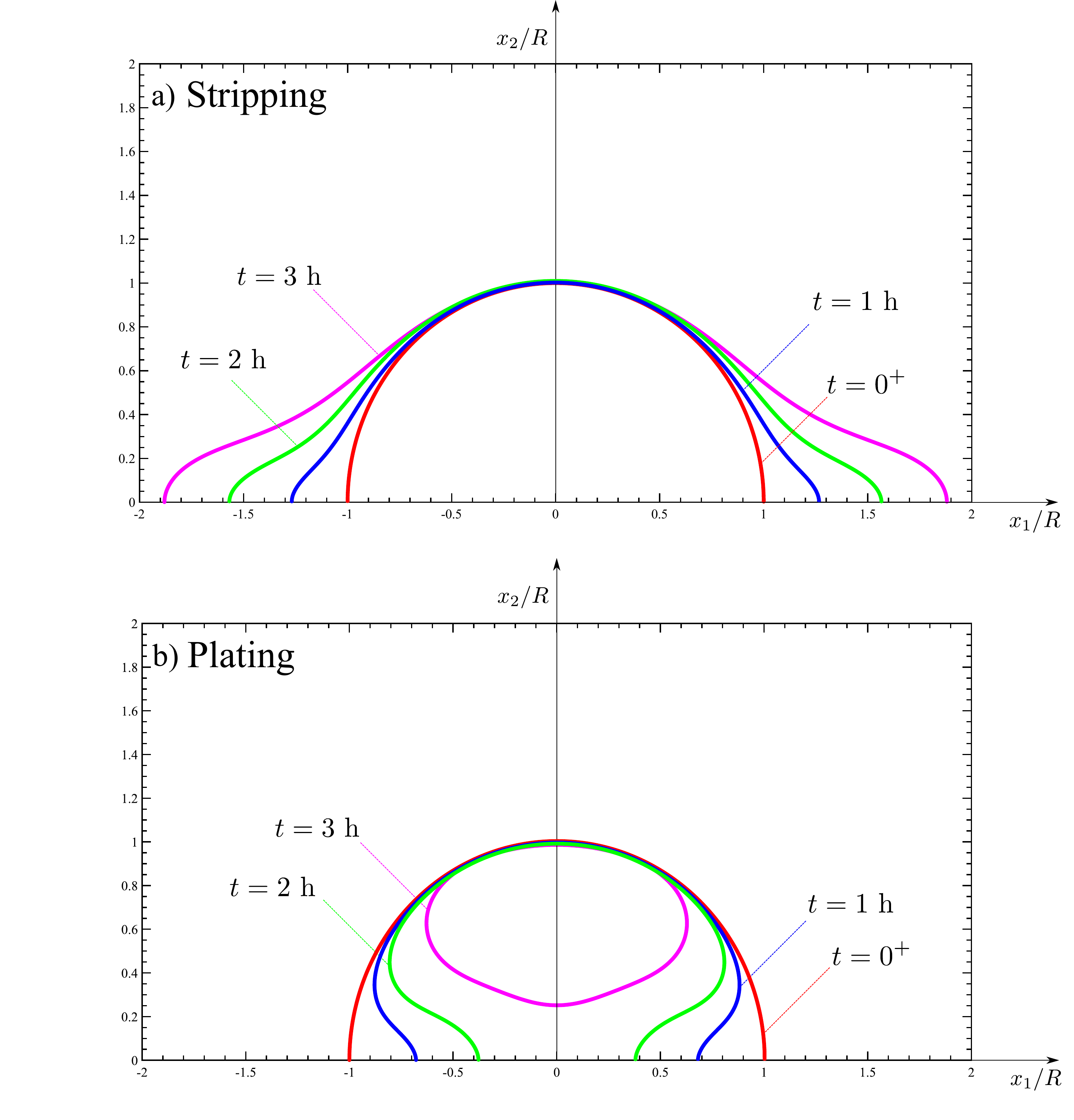}
    \caption{Numerical experiments on a single void model: quantifying changes in void shape for (a) stripping and (b) plating. The results are shown relative to a Cartesian coordinate system ($x_1,x_2$) whose origin is located in the centre of the circle containing the semi-circular void, with $x_2$ being perpendicular to the electrode-electrolyte interface.}
    \label{fig:visco_plating_stripping_compare}
\end{figure}

On the other hand, see Fig. \ref{fig:single_void_visco_plating_contour}, the results obtained for plating show the opposite trend. The opening of the void narrows at the electrode-electrolyte interface and eventually closes completely, isolating the void from the electrolyte. Thus, see Fig. \ref{fig:introduction}, the model captures the two phenomena that can be observed during plating: the reduction in void size and void occlusion. It is worth noting that the regions of high current reduce in size as the plating process evolves, and that these hot spots eventually disappear, recovering an intact electrode-electrolyte interface. The exact void shapes are reported as a function of time in Fig. \ref{fig:visco_plating_stripping_compare}, for both stripping and plating. Here, the void-electrode interface is taken to be described by the $\xi=0.5$ iso-contour. It can be seen how the void changes shape mostly near the electrode-electrolyte interface, contracting during plating and expanding during stripping. At the end of the stripping cycle, the length of the region where electrode and electrolyte are no longer in contact has almost duplicated. 

\subsubsection{Influence of the applied current and the phase field kinetic coefficient}
\label{Sec:LANDCurrentDensity}

We proceed to conduct a parametric analysis to investigate the influence of the applied current density $i_\mathrm{app}$ and the phase field kinetic parameter $L$. The aim is to investigate the competition between relevant kinetic phenomena. Thus, voiding is governed by four kinetic events: the rate of Li dissolution/deposition, bulk Li transport, the nucleation/annihilation of vacancies, and creep. Let us momentarily leave aside the role of creep, which is investigated below (Section \ref{Sec:CreepVSdiffusion}). The first three events are governed by the applied current $i_\mathrm{app}$, the effective diffusion coefficient $D_{\mathrm{eff}}$, and the phase field mobility coefficient $L$, respectively. Since $D_{\mathrm{eff}}$ is known and can be independently measured, we focus our attention on the sensitivity to $i_\mathrm{app}$ and $L$. The boundary value problem corresponds to that considered in the previous section and depicted in Fig. \ref{fig:SingleVoidConfig}. Again, no pressure is applied so as to isolate effects. The outcome of the simulations is reported in terms of void shape evolution for selected values of $i_\mathrm{app}$ and $L$ at a time of $t=1$ h. As in Fig. \ref{fig:visco_plating_stripping_compare}, the results are shown in a normalised Cartesian coordinate system ($x_1/R,x_2/R$) whose origin is located in the centre of the circle containing the semi-circular void, with $x_2$ being perpendicular to the electrode-electrolyte interface.\\

\begin{figure}[H]
    \centering
    \includegraphics[width=0.8\linewidth]{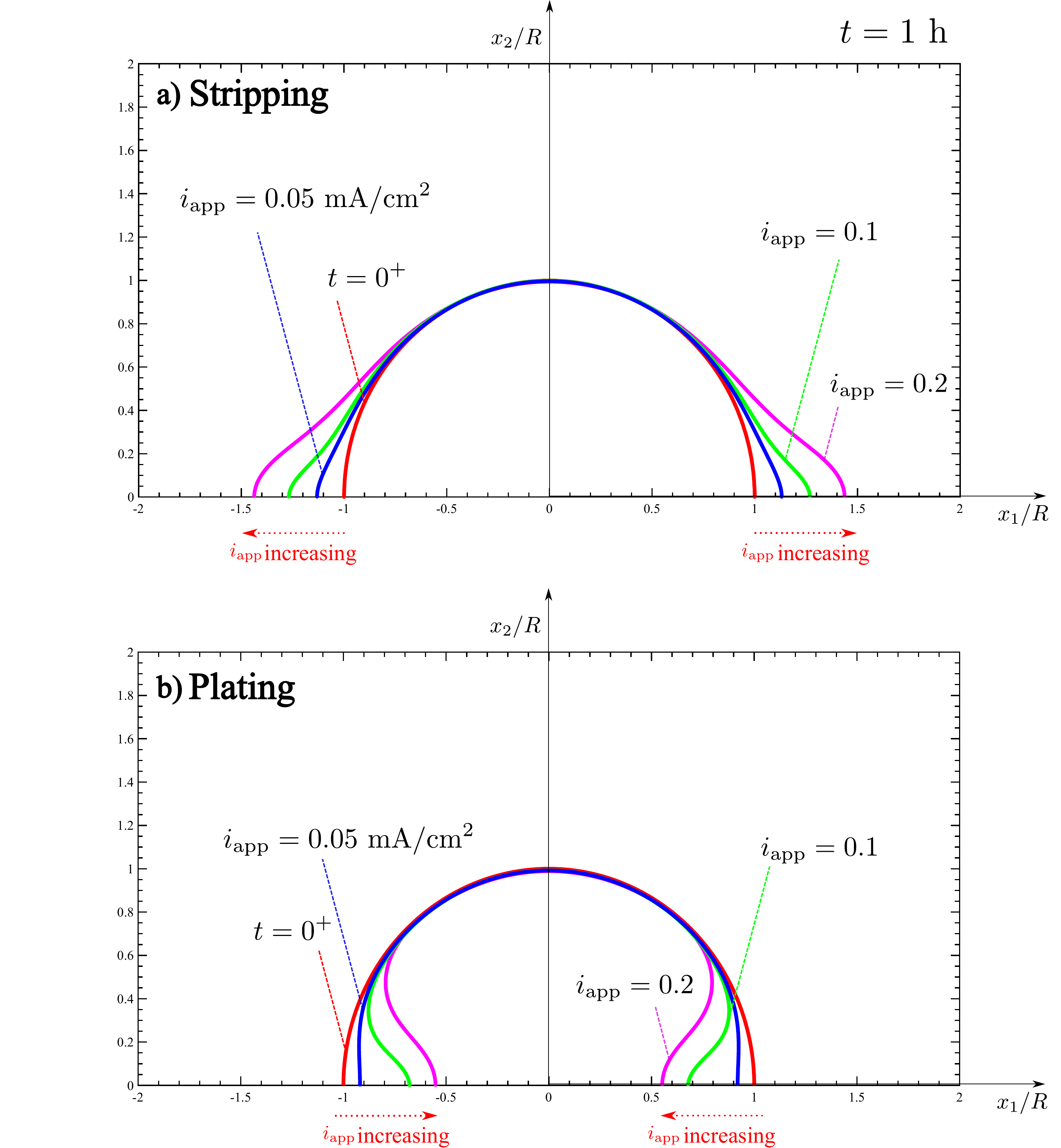}
    \caption{Numerical experiments on a single void model: quantifying the role of the applied current $i_{\mathrm{app}}$ for (a) stripping and (b) plating. Results are shown for a time of \SI{1}{\hour} and relative to a Cartesian coordinate system ($x_1,x_2$) whose origin is located in the centre of the circle containing the semi-circular void, with $x_2$ being perpendicular to the electrode-electrolyte interface.}
    \label{fig:i0_sweep_stripping_and_plating_compare}
\end{figure}

The results obtained for different values of the applied current density are given in Fig. \ref{fig:i0_sweep_stripping_and_plating_compare}. In agreement with expectations and experimental observations, the sensitivity of the void shape to plating and stripping increases with increasing $i_\mathrm{app}$. By comparing with Fig. \ref{fig:visco_plating_stripping_compare}, we can see that the void evolution appears to be qualitatively very similar for all the choices of $i_\mathrm{app}$ considered, with the main effect being related to the rate at which these changes take place. Thus, the ratio of $i_\mathrm{app}$ to the charging cycle time is a key factor. For example, the generation of voids will be minimised if the stripping current is low relative to the duration of the stripping cycle. Moreover, voiding is very sensitive to the interplay between $i_\mathrm{app}$ and $D_{\mathrm{eff}}$, as the formation of voids will be entirely suppressed if the stripping current removes Li from the interface more slowly than it can be replenished. For the values of $i_\mathrm{app}$ considered, this is not observed in the stripping results reported in Fig. \ref{fig:i0_sweep_stripping_and_plating_compare}. However, simulations conducted with smaller values of $i_\mathrm{app}$ reveal that no noticeable void growth takes place after $t=\SI{1}{\hour}$ when the applied current is equal to 0.005 \si{\milli\ampere\per\centi\meter\squared} or smaller. Thus, our simulations are consistent with the existence of a critical current density, as inferred experimentally \citep{Kasemchainan2019}.\\

\begin{figure}[H]
    \centering
    \includegraphics[width=0.8\linewidth]{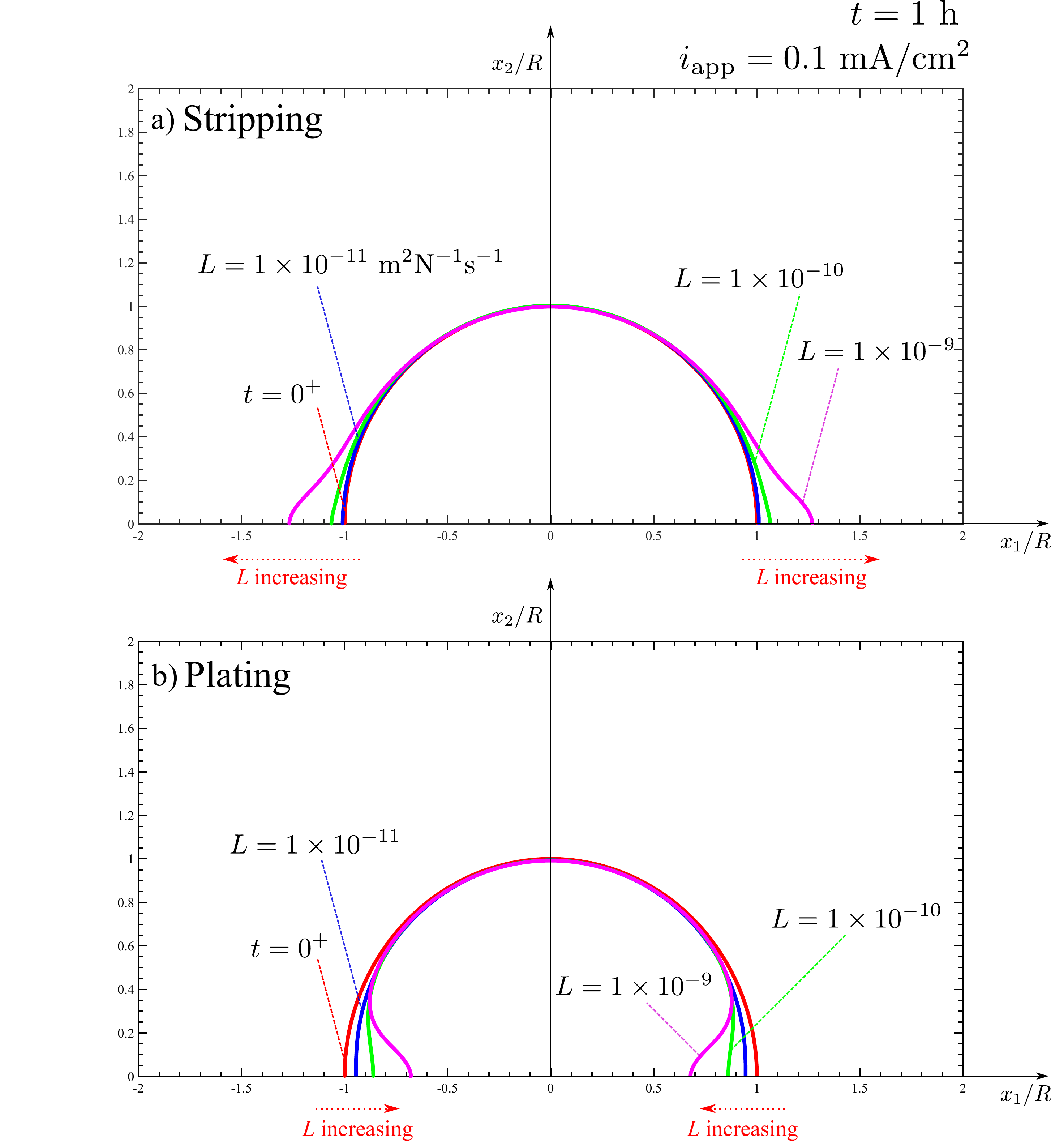}
    \caption{Numerical experiments on a single void model: quantifying the role of the phase field kinetic parameter $L$ for (a) stripping and (b) plating. Results are shown for a time of \SI{1}{\hour} and relative to a Cartesian coordinate system ($x_1,x_2$) whose origin is located in the centre of the circle containing the semi-circular void, with $x_2$ being perpendicular to the electrode-electrolyte interface.}
    \label{fig:L_sweep_stripping_and_plating_compare}
\end{figure}

We shall now investigate the role of the phase field kinetic parameter $L$, also termed phase field mobility coefficient. In this formulation, $L$ characterises the ability of a material to nucleate or annihilate vacancy sites. The results obtained for selected values of $L$ are shown in Fig. \ref{fig:L_sweep_stripping_and_plating_compare}, for both stripping and plating processes. The stripping results reveal that low $L$ values ($L\sim\num{e-11}$) result in a void that largely maintains its original shape. However, the change in shape is noticeable for values of $L$ on the order of $\sim\num{e-10}$ or larger, for the $t=\SI{1}{\hour}$ of stripping considered. As in the previous stripping simulations, void growth is mostly enhanced near the interface, smearing the void and losing the initial semi-circumferential shape. During plating, the void shrinks inwards from all sides for small $L$ values, while large $L$ values enhance the closing of the void opening at the electrode-electrolyte interface. If $L$, $i_\mathrm{app}$ or the plating time are sufficiently large, then the void opening closes completely and the void becomes occluded in the electrolyte (see Fig. \ref{fig:visco_plating_stripping_compare}b). These simulations suggest that $L$ could be obtained from first principles calculations or inferred from experiments by applying inverse engineering. Further insight into the interface velocity is gained analytically in \ref{subsec:interface_velocity}. 

\subsubsection{Investigating the competing role of creep deformation and Li diffusion}
\label{Sec:CreepVSdiffusion}

\begin{figure}[H]
    \centering
    \includegraphics[width=0.8\linewidth]{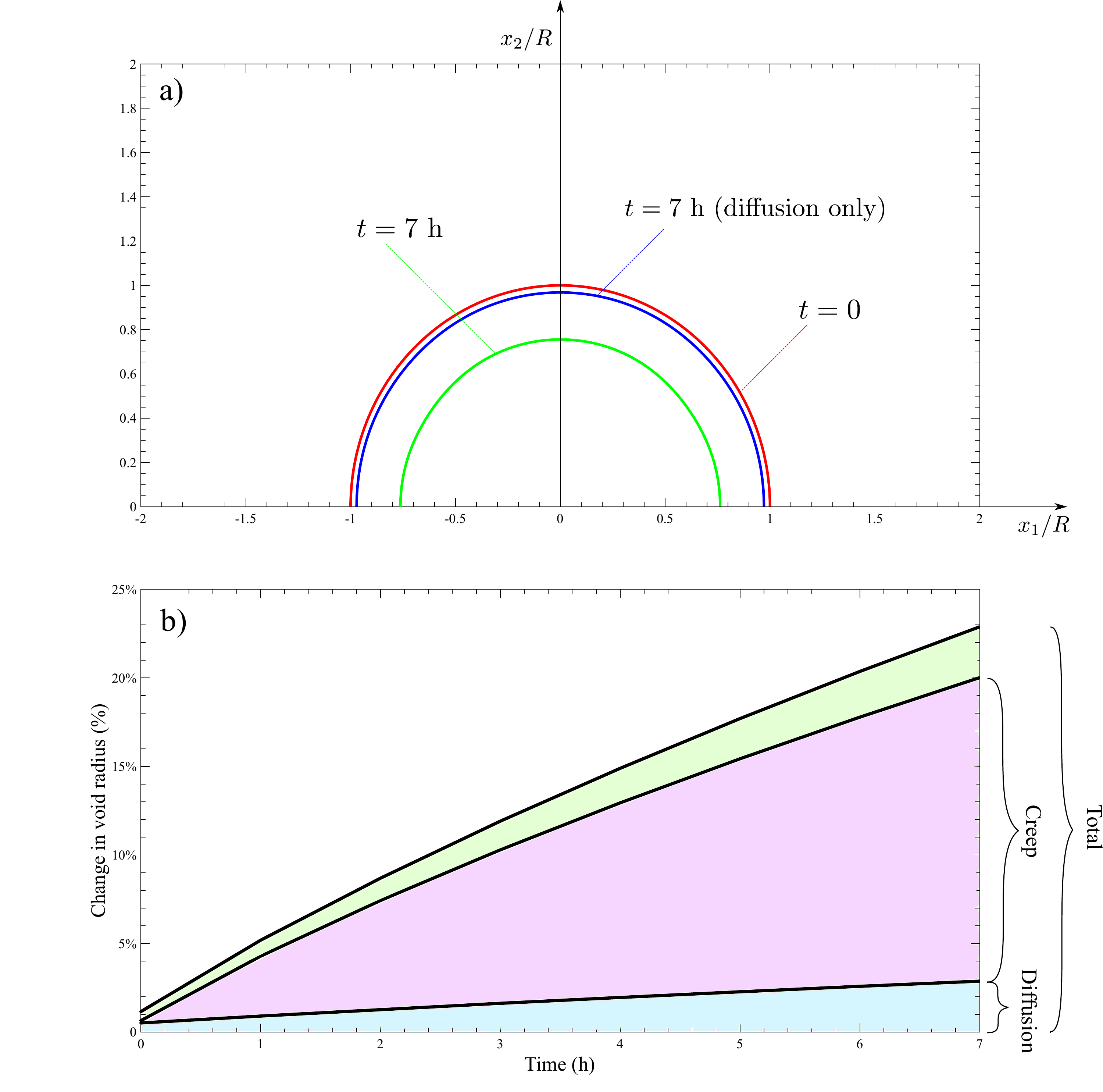}
    \caption{Numerical experiments on a single void model: quantifying the competition between Li diffusion and creep; (a) void shape evolution relative to a Cartesian coordinate system ($x_1,x_2$) whose origin is located in the centre of the circle containing the semi-circular void, with $x_2$ being perpendicular to the electrode-electrolyte interface, and (b) percentage change in void radius versus time.}
    \label{fig:creep_effect_compare}
\end{figure}

Finally, we use the single void boundary value problem to gain insight into the competition between diffusion and creep. We isolate these effects by assuming that no flux is applied on the electrode-electrolyte interface and by replacing the electrolyte and its mechanical constraint by a fixed displacement $u_x = 0$ at $\Gamma^f$. A pressure with magnitude $p=\SI{0.6}{\mega\pascal}$ is applied on the left edge of the electrode. This pressure is applied instantaneously (time $t=0^+$) and then held fixed for \SI{7}{\hour}, such that creep strains significantly out-weight their elastic counterparts. The results obtained are shown in Fig. \ref{fig:creep_effect_compare} for three scenarios. Consider first the void shape evolution results, Fig. \ref{fig:creep_effect_compare}a. Three curves are shown to characterise the contributions of Li diffusion and creep to void shape evolution: (i) the initial void shape, before the pressure is applied (time $t=0$); (ii) the void shape after \SI{8}{\hour} if the undeformed shape is considered (i.e., the change in shape due to diffusion only); and (iii) the void shape after \SI{8}{\hour} considering both diffusion and material deformation contributions. The results reveal that, in the absence of current, the void shrinks while maintaining a semi-circular shape --- no localisation events are observed. It is also seen that the change in void shape due to diffusion appears to be a small contribution to the final void shape. 

Quantitative insight can be gained from Fig. \ref{fig:creep_effect_compare}b, where the contributions from creep and Li diffusion to void evolution are shown in terms of the relative change of the void radius. It can be seen that creep deformation dominates the voiding process, relative to the contribution from Li diffusion. This is in agreement with recent experimental measurements for various levels of applied pressure, which suggest that creep rather than diffusion dominates the rate at which Li is replenished at the interface \citep{Kasemchainan2019}. The precise weighting of the contributions from diffusion and creep will depend on the total time and the material's ability to annihilate vacancies.  

\subsection{Cyclic charging of a solid-state cell with multiple interface defects}
\label{Sec:FullModel}

\begin{figure}[H]
    \centering
    \includegraphics[width=1\linewidth]{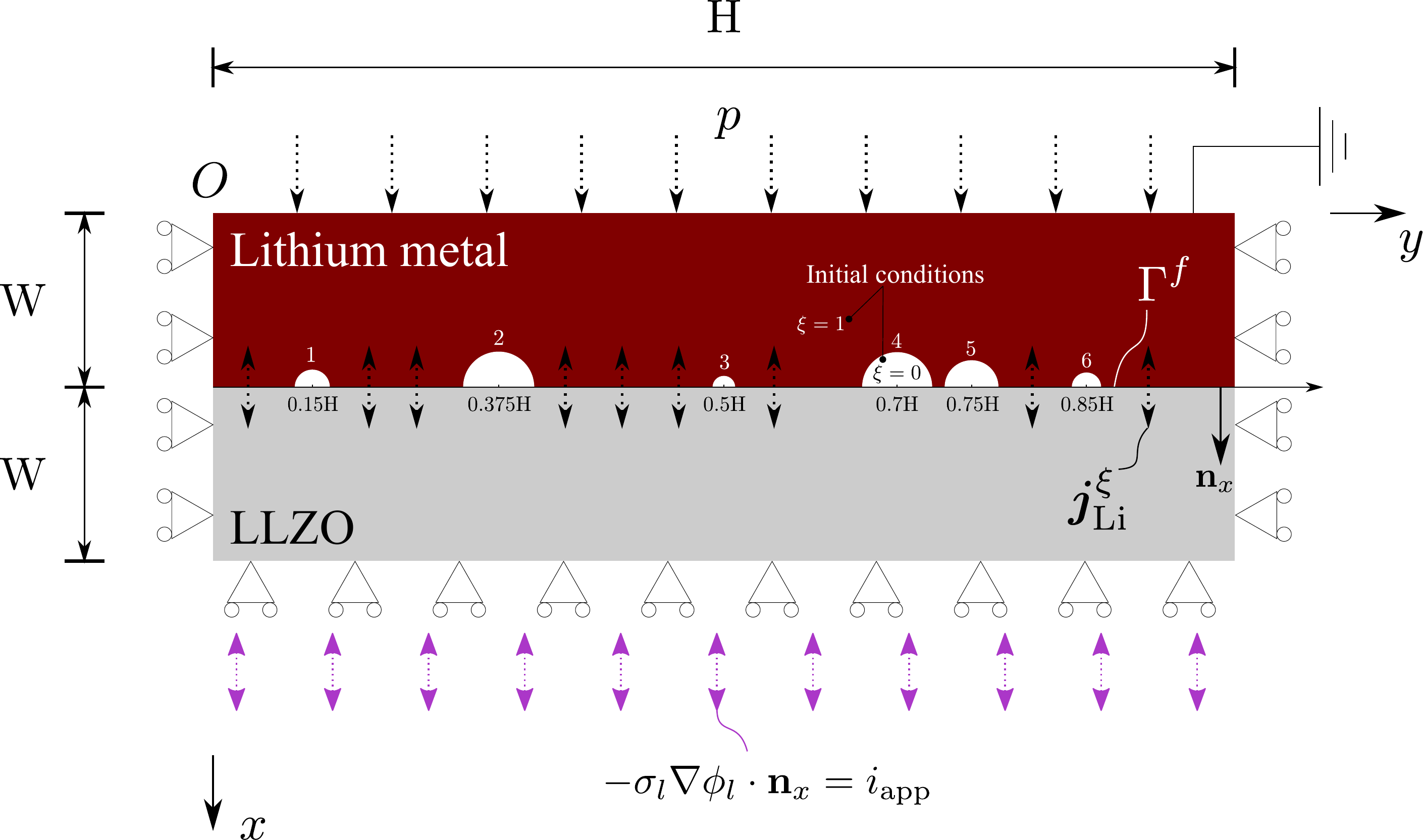}
    \caption{Solid-state cell with multiple interface defects: dimensions, configuration and boundary conditions. Initial and boundary conditions are also provided. Six voids of different radii are introduced to induce an uneven current distribution; as quantified in the schematic, these voids are placed at arbitrary locations along the electrode-electrolyte interface.}
    \label{fig:visco_multiple_void_with_pressure_setting}
\end{figure}

We proceed now to predict the evolution of voids and hot-spots in an all-solid-state cell undergoing multiple plating and stripping cycles. The aim is two-fold: to gain new insight and to showcase the abilities of the model in delivering predictions under realistic conditions. The schematic of the boundary value problem is given in Fig. \ref{fig:visco_multiple_void_with_pressure_setting}. The electrode-electrolyte system has dimensions of $\mathrm{H}=\SI{250}{\micro\meter}$ and $\mathrm{W}=0.16\mathrm{H}$. The current at the electrode-electrolyte interface is unevenly distributed because of the presence of defects and surface roughness. This is captured by defining six small voids that have different radii and are located in arbitrary positions along the anode-electrolyte interface. These voids, numbered from left to right in Fig. \ref{fig:visco_multiple_void_with_pressure_setting}, have radii of $\mathrm{R}_1=0.0132\mathrm{H}$, $\mathrm{R}_2=0.025\mathrm{H}$, $\mathrm{R}_3=0.0084\mathrm{H}$, $\mathrm{R}_4=0.02\mathrm{H}$, $\mathrm{R}_5=0.0168\mathrm{H}$, and $\mathrm{R}_6=0.0112\mathrm{H}$. 
Their precise positions are provided in Fig. \ref{fig:visco_multiple_void_with_pressure_setting} through the relative location of the centre of the circles containing each of the semi-circular voids. The initial and boundary conditions of the problem are also given in Fig. \ref{fig:visco_multiple_void_with_pressure_setting}; these aim at mimicking the operating conditions of solid-state cells. From the point of view of diffusion, a flux is defined at the electrode-electrolyte interface, whose magnitude is determined by the current density. Regarding the electrical problem, a uniform current density $i_\mathrm{app}$ is prescribed in the free surface of the LLZO electrolyte. Mechanically, we apply a pressure $p$ to the free surface of the Li metal anode but also evaluate its influence by considering the case of $p=0$, where the normal displacement component is constrained.

\begin{figure}[H]
    \centering
    \includegraphics[width=0.9\linewidth]{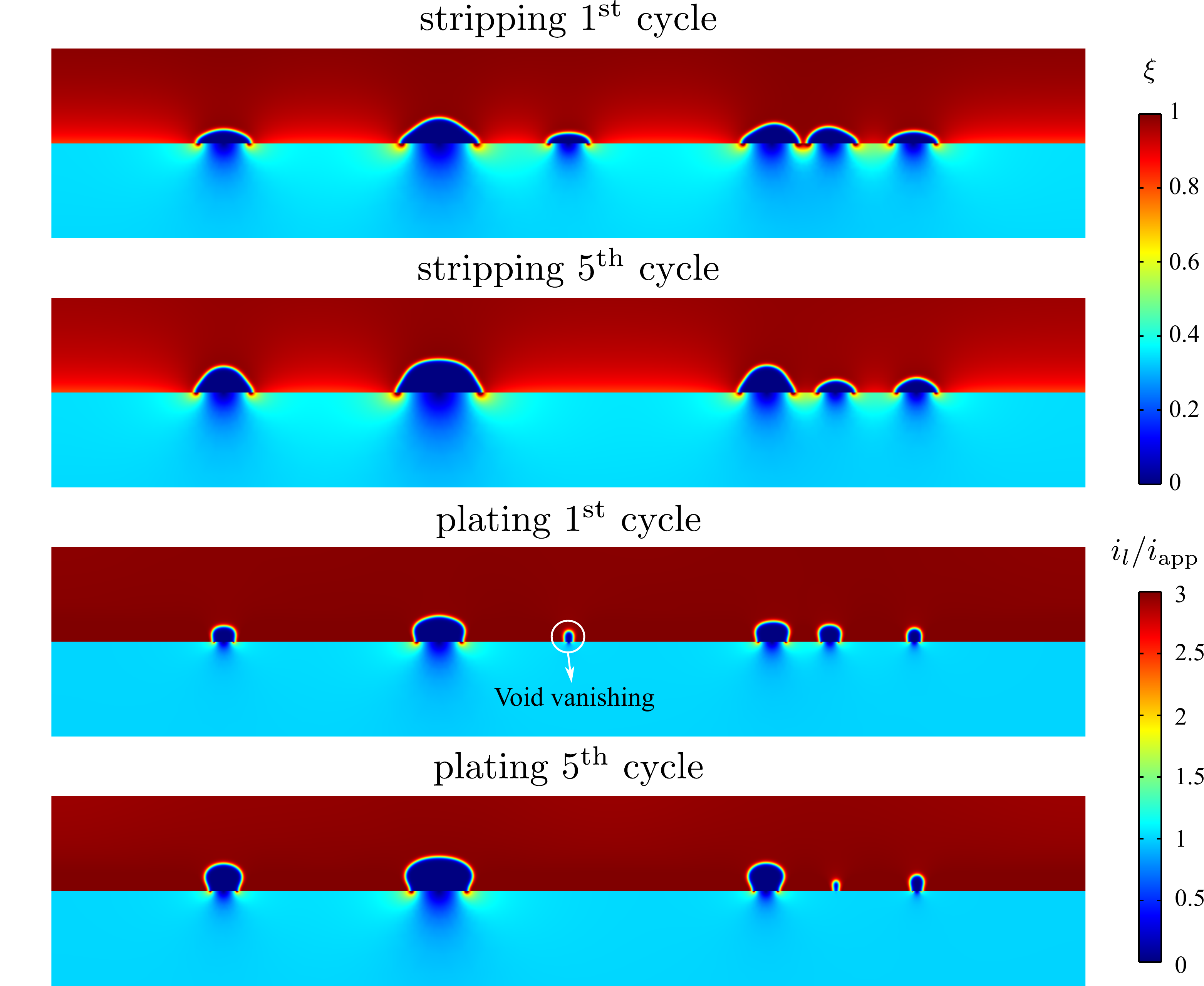}
    \caption{Numerical experiments on a Li metal anode - LLZO electrolyte system under realistic conditions: predictions of void evolution and current distribution for several stripping and plating cycles.}
    \label{fig:CellMultipleCycles}
\end{figure}

First, we investigate the evolution of the voids' morphology during multiple cycles of plating and stripping. Specifically, we simulate 5 charging cycles, each of which lasts for \SI{2.5}{\hour}; stripping regimes of  \SI{1.25}{\hour} followed by plating periods of the same duration. The results obtained in the absence of an applied pressure ($p=0$) are shown in Fig. \ref{fig:CellMultipleCycles}. Several interesting features are observed. In the first stripping regime, we find that all voids enlarge along the interface, with current hot spots appearing at their edges. The voids, initially semi-circular, widen and notably change their aspect ratio, going towards an elliptical shape. This effect is more pronounced for the smaller voids (voids 1, 3 and 6). If voids are close to each other (see voids 4 and 5), the symmetry of their shape is lost and coalescence is observed. After \SI{1.25}{\hour}, the current is inverted and plating takes place. As observed for the single void analysis (see Fig. \ref{fig:single_void_visco_plating_contour}), the edges of each void become closer to each other. However, unlike the single void study, the void edges do not come into contact in the first plating cycle, as the initial state for plating is the stripped electrode. 

As the charging and discharging cycles proceed, the majority of the voids increase their size; see, e.g., the evolution of void 2 between the first and fifth plating cycles. However, the smallest void (void 3) vanishes during the fourth plating period and therefore current hot-spots are not subsequently present in that region. We also observe how the void shape changes with plating and stripping cycles, adopting aspect ratios closer to the initial one. Overall, we see a deterioration of the electrolyte-electrode interface, with voids increasing in size, and a reduction in the electrolyte-electrode contact area.\\

\begin{figure}[H]
    \centering
    \includegraphics[width=\linewidth]{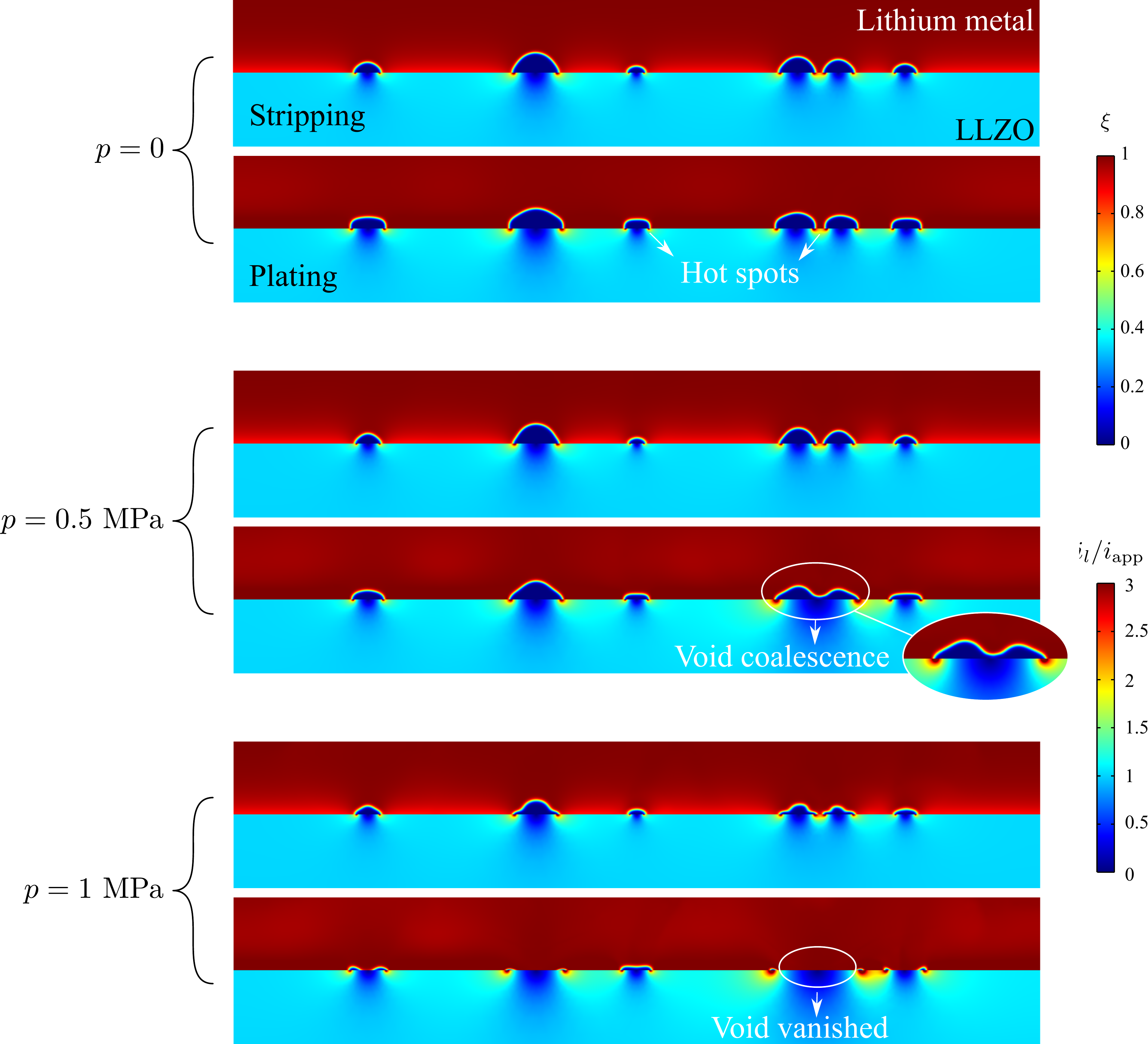}
    \caption{Numerical experiments on a Li metal anode - LLZO electrolyte system under realistic conditions: influence of the applied pressure $p$ in the void evolution and current distribution. A complete charge-discharge cycle over 2.5 h is simulated, starting with a 1.25 h stripping half cycle. The figures display the results obtained after 30\% of the stripping ($t=22.5$ min.) and plating ($t=97.5$ min.) half cycles.}
    \label{fig:multiple_cyclic_pressure vs no pressure}
\end{figure}

It has been frequently argued that the most promising strategy to prevent the interfacial instabilities that result from voiding is the application of mechanical pressure \citep{Sakamoto2019,Kasemchainan2019}. Small stripping currents can readily lead to rates of Li oxidation that are greater than the rate at which Li is replenished due to bulk Li diffusion. Hence, the aim is to apply a constant pressure to provide another source of Li to the interface \textit{via} creep deformation. As shown in Section \ref{Sec:CreepVSdiffusion}, creep plays a much more significant role in shrinking voids than Li diffusion and thus is key to minimising voiding and the emergence of current hot spots. We proceed to apply a constant pressure during plating and stripping, and compare the outcome of the simulations with the predictions obtained for $p=0$. The results obtained for selected values of the applied pressure ($p=0$, 0.5 and \SI{1}{\mega\pascal}) are shown in Fig. \ref{fig:multiple_cyclic_pressure vs no pressure}. 

It can be observed that the application of pressure is very effective in reducing the size of the voids. The cases with $p > 0$ show ellipsoidal voids that flatten with time, up to the point of vanishing for $p=\SI{1}{\mega\pascal}$. This leads to the suppression of current hot spots that are otherwise present at the void ligaments (see the region beetween voids 4 and 5 for $p=0$). Significant differences are observed with the case of no pressure, in terms of the electrode-electrolyte contact area. Phenomena such as void coalescence can also be observed for intermediate pressures ($p=\SI{0.5}{\mega\pascal}$). The sensitivity of void size to pressure is important as smaller voids lead to a higher conductance of the battery cell. For the charging conditions and materials considered here, a constant pressure of \SI{1}{\mega\pascal} appears to be sufficient to significantly reduce voiding.

\section{Summary and concluding remarks}
\label{Sec:Conclusions}

We have presented a phase field-based electro-chemo-mechanical model that can predict the evolution of voids and current hot-spots as a function of material properties, applied current and mechanical pressure. To capture the phenomena governing voiding and other interfacial instabilities in all-solid-state battery cells, the theory combines substitutional Li diffusion, a phase field description of vacancy nucleation and annihilation, and a viscoplastic constitutive model for Li metal. Moreover, the framework is extended to predict deformation and current distribution in the electrolyte, accounting for electrode-electrolyte interactions and capturing the emergence of regions of high current near the void edges. Being able to predict the nucleation and intensity of these local current `hot spots' is of utmost importance as they lead to the formation of dendrites and subsequent cell short-circuit. To deliver predictions over relevant time and space scales, the model is numerically implemented using the finite element method. Case studies involving stripping and plating cycles in electrode-electrolyte systems with one or multiple voids are carried out to showcase the predictive capabilities of the model and gain insight into interfacial stabilities in all-solid-state batteries. Calculations are conducted to investigate the competition and interplay between creep, bulk and surface Li diffusion, Li dissolution and deposition, and vacancy nucleation and annihilation. Our main findings are:
\begin{itemize}
\item Void morphology is very sensitive to charging conditions. As observed experimentally, stripping currents lead to an expansion of the void along the interface, while plating translates into a contraction of the void edges, potentially leading to the complete void occlusion. These phenomena are magnified with increasing applied current and in materials that can more readily annihilate vacancies (as characterised by the phase field kinetic parameter $L$).
\item High magnitudes of current density are predicted near the void edges, with the region where local current is high ($i_l/i_\mathrm{app}>3$) increasing in size with the stripping time.
\item The application of a mechanical pressure on the free surface of the electrode notably reduces voiding and leads to much larger contact areas between electrode and electrolyte. We observe that smaller voids can eventually vanish under relevant conditions if the pressure is sufficiently large.
\item Creep is found to play a much more significant role than bulk Li diffusion in governing the shape of the voids. 
\item Complex voiding phenomena such as void coalescence are found to govern interfacial instabilities when simulating realistic conditions (multiple voids and charging cycles).
\end{itemize}

This work can be further enhanced by incorporating a dendrite nucleation criterion or explicitly simulating dendrite evolution. Thus, the theoretical and computational framework presented establishes the basis for the development of advanced prognosis tools that can accelerate future developments in battery technology.

\section{Acknowledgements}
\label{Sec:Acknowledgeoffunding}

Ying Zhao acknowledges financial support from the National Natural Science Foundation of China [Project No. 12102305], Shanghai Sailing Program [Project No. 20YF1452300] and the Fundamental Research Funds for the Central Universities. Emilio Mart\'{\i}nez-Pa\~neda acknowledges financial support from UKRI's Future Leaders Fellowship programme [grant MR/V024124/1].

\appendix

\setcounter{figure}{0}

\section{Interface velocity}
\label{subsec:interface_velocity}

Let us gain analytical insight into the nature of the interface velocity by simplifying the model presented in Section \ref{Sec:Theory}. Following experimental observations \citep{Kasemchainan2019} and the results obtained in Section \ref{Sec:CreepVSdiffusion}, we proceed to assume that bulk Li diffusion plays a secondary role in the evolution of the void-anode interface (i.e., $\theta_\m$ varies less significantly than $\xi$). Upon adopting this assumption, one can set $\theta_\m=\theta_\m^0$ and simplifiy the governing equation (\ref{eq:govern_diffusion}) to
\begin{align}
\frac{\partial h}{\partial t} = -\frac{D_\mathrm{eff}(\Omega_\Li-\Omega_\vac)}{RT}\bm\nabla \cdot h\bm\nabla\sigma_h^\xi
\end{align}
which can be re-arranged as,
\begin{align}
\frac{\partial h}{\partial t} = -\frac{D_\mathrm{eff}h(\Omega_\Li-\Omega_\vac)}{RT}\bm\nabla^2\sigma_h^\xi-\frac{D_\mathrm{eff}(\Omega_\Li-\Omega_\vac)}{RT}\bm\nabla h\cdot\bm\nabla\sigma_h^\xi.\label{eq:govern_diffusion_middle}
\end{align}

It is anticipated that there are no sinks or sources of hydrostatic stresses in the Li metal anode (e.g. point loads or atom sinks), which leads to $\bm\nabla^2\sigma_h^\xi=0$ in the bulk ($h=1$). Thus, by defining $\theta_\m$ as constant, diffusion is constrained to the interface between the void and the lithium metal.
%
%
We denote $s$ as the moving front displacement of the interface, whose direction is normal to the interface and can be expressed by $\bm\nabla h/\left|\bm\nabla h\right|$. The tangential direction is denoted by $u$. Recalling that
\begin{align}
\frac{\partial \left(\cdot\right)}{\partial s} = \frac{\bm\nabla h}{\left|\bm\nabla h\right|}\cdot\bm\nabla\left(\cdot\right)
\end{align}
where $\left(\cdot\right)$ represents a field variable of interest, Eq. (\ref{eq:govern_diffusion_middle}) can be expressed as
\begin{align}
\frac{\partial s}{\partial t} = -\frac{h}{\left|\bm\nabla h\right|}\frac{D_\mathrm{eff}(\Omega_\Li-\Omega_\vac)}{RT}\left(\frac{\partial^2\sigma_h^\xi}{\partial s^2} +\frac{\partial^2\sigma_h^\xi}{\partial u^2} \right) -\frac{D_\mathrm{eff}(\Omega_\Li-\Omega_\vac)}{RT}\frac{\partial\sigma_h^\xi}{\partial s}.
\label{eq:interface_velocity}
\end{align}
In quasi-static problems, the pressure is equilibrated by surface tension, which means that both $\partial\sigma_h^\xi/\partial s$ and $\partial^2\sigma_h^\xi/\partial s^2$ vanish in the surface. Thus, Eq. (\ref{eq:interface_velocity}) can be further simplified to
\begin{align}
\frac{\partial s}{\partial t} = -\frac{h}{\left|\bm\nabla h\right|}\frac{D_\mathrm{eff}(\Omega_\Li-\Omega_\vac)}{RT}\frac{\partial^2\sigma_h^\xi}{\partial u^2},\label{eq:interface_velocity_simplified}
\end{align}
which has similar form as the governing equation for surface diffusion; see, e.g.,  \citet{Chuang1979,NEEDLEMAN1983}. The regulation term ${h}/{\left|\bm\nabla h\right|}$ has a dimension of length, representing the width of the diffusion layer. The source of hydrostatic stress in the interface is the nucleation or annihilation of lattice sites, which is governed by the phase field equation (\ref{eq:govern_evolution_xi}). Accordingly, Eq. (\ref{eq:govern_evolution_xi}) is simplified to
\begin{align}
\frac{\partial\xi}{\partial t} =  - L\Omega_\vac c_\Lat^\m h^\prime\psi_e - Lwg^\prime + L\kappa\nabla^2\xi \, , \label{eq:govern_evolution_xi_simplified}
\end{align}

\noindent which clearly shows that, when bulk diffusion is neglected, vacancy nucleation and annihilation can only take place in the interface ($0<\xi<1$).

\section{Interfacial energy and thickness based on the phase field model}
\label{App:interface_thickness}

In the absence of mechanical and chemical contributions, the interfacial Helmholtz free energy density across the interface reads
\begin{align}
\Psi^\inter = A^\inter\int_{-\infty}^{+\infty} w\xi^2\left(1-\xi\right)^2+\frac{1}{2}\kappa\left(\frac{\mathrm{d}\xi}{\mathrm{d}s}\right)^2\,\mathrm{d}s\label{eq:1D_interfacial_energy}
\end{align}
where $A^\inter$ denotes the area of the interface and $s$ is in the direction normal to the interface. In equilibrium $\delta\Psi^\inter=0$ and the following Euler equation holds:
\begin{align}
I-\left(\frac{\mathrm{d}\xi}{\mathrm{d}s}\right)\left[\frac{\partial I}{\partial\left(\mathrm{d}\xi/\mathrm{d}s\right)}\right]=\mathrm{const.} \label{eq:Euler_equation}
\end{align}
where $I$ is the integrand of Eq. (\ref{eq:1D_interfacial_energy}). From Eqs. (\ref{eq:Euler_equation})-(\ref{eq:1D_interfacial_energy}) one reaches
\begin{align}
w\xi^2\left(1-\xi\right)^2-\frac{1}{2}\kappa\left(\frac{\mathrm{d}\xi}{\mathrm{d}s}\right)^2=\mathrm{const.}\label{eq:Eular_equation_phase field}
\end{align}
for all $s\in\left(-\infty,+\infty\right)$.
We assume that $\xi=0$ in the limit of $s\to-\infty$ and $\xi=1$ in the limit of $s\to+\infty$. Then, it is clear that ${\mathrm{d}\xi}/{\mathrm{d}s}\ge0$ and that Eq. (\ref{eq:Eular_equation_phase field}) yields
\begin{align}
\frac{\mathrm{d}\xi}{\mathrm{d}s}=\sqrt{\frac{2w}{\kappa}}\xi\left(1-\xi\right).\label{eq:dxids}
\end{align}
If we then define the location of the interface at $\xi=0.5$ ($s=s_0$), the solution for $\xi$ reads
\begin{align}
\xi = \frac{1}{\exp\left[-\sqrt{\frac{2w}{\kappa}}\left(s-s_0\right)\right]+1}.
\end{align}
The interfacial thickness $\ell$ and energy $\Psi^\inter/A^\inter$ are then respectively derived as
\begin{align}
&\ell= \left.\frac{1}{\mathrm{d}\xi/\mathrm{d}s} \right\rvert_{s=s_0}=\sqrt{\frac{8\kappa}{w}}\quad\mathrm{and}\label{eq:expression_interfacial_thickness}\\
&\Psi^\inter/A^\inter= \int_0^12w\xi^2\left(1-\xi\right)^2 \sqrt{\frac{\kappa}{2w}}\frac{1}{\xi\left(1-\xi\right)}\,\mathrm{d}\xi=\frac{\sqrt{2\kappa w}}{6} \, .\label{eq:expression_interfacial_energy}
\end{align}
It is anticipated that the interfacial energy density (or surface tension) can be estimated based on Eqs. (\ref{eq:expression_interfacial_thickness}) and (\ref{eq:expression_interfacial_energy}) as
\begin{align}
 \gamma = \frac{\Psi^\inter}{\ell A^\inter}=\frac{w}{12}.
 \end{align} 

\begin{figure}[H]
    \centering
    \includegraphics[width=0.8\linewidth]{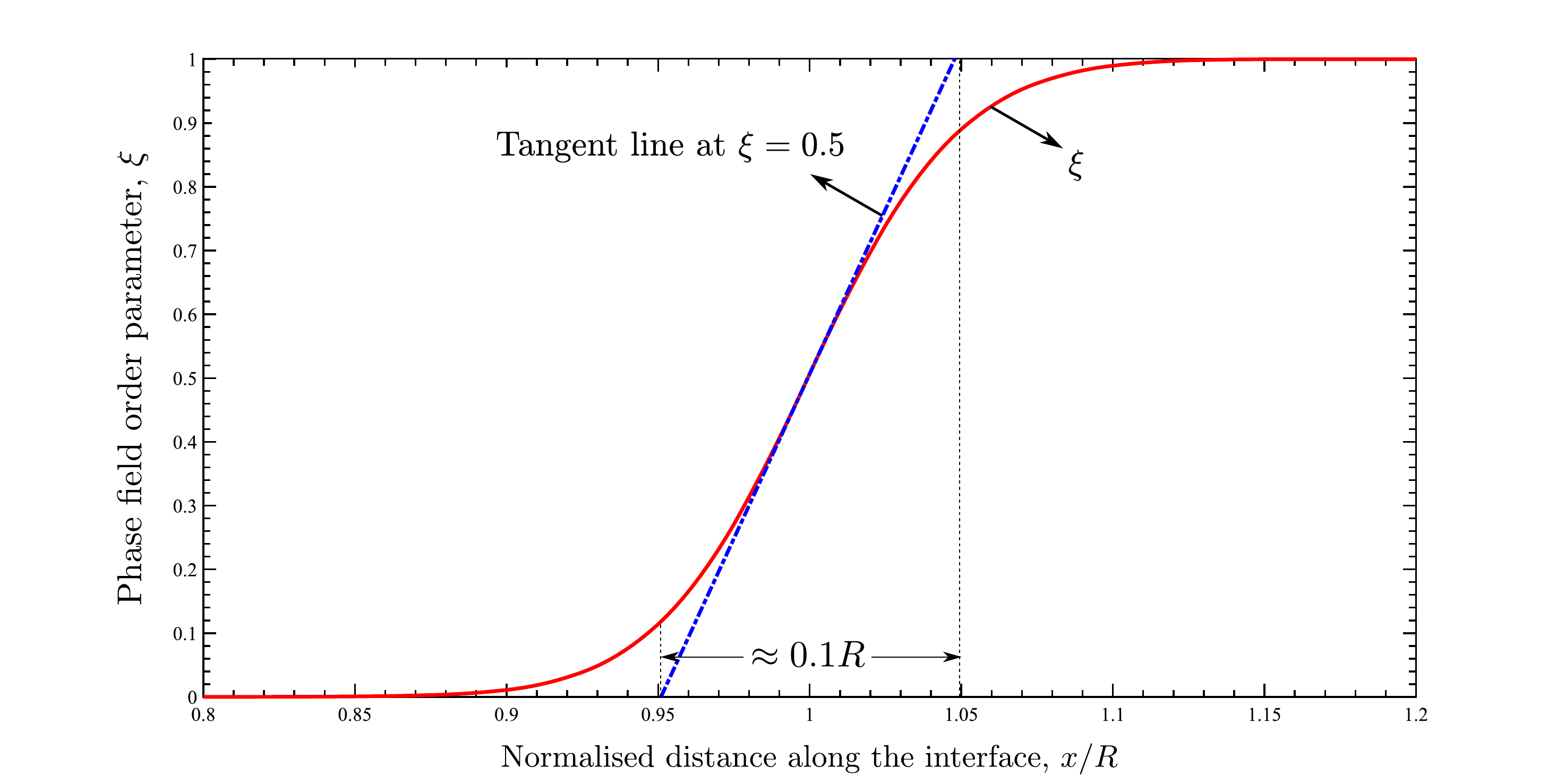}
    \caption{Phase field interface thickness: comparison between the theoretical estimate ($\ell=1$ \si{\micro\meter}, for $\kappa =\SI{4.5e-7}{\newton} $ and $w=\SI{3.5e6}{\newton\per\meter\squared}$) and the numerical prediction. The distance along the interface is normalised by the void radius ($\mathrm{R}=10$ \si{\micro\meter}). For consistency, the interface thickness is determined by drawing a line tangent to the the $\xi$ distribution predicted at $\xi=0.5$.}
    \label{fig:interface_thickness}
\end{figure}

We shall now compare this analytical estimate with the predictions from the finite element model. To achieve this, we simulate the single void boundary value problem depicted in Fig. \ref{fig:SingleVoidConfig}. No pressure or flux is applied and the simulation is run until the equilibrium state is reached. The predicted $\xi$ distribution for the choices of $\kappa =\SI{4.5e-7}{\newton} $ and $w=\SI{3.5e6}{\newton\per\meter\squared}$ is shown in Fig. \ref{fig:interface_thickness}, where the distance along the interface is shown normalised by the void radius (with $\mathrm{R}=10$ \si{\micro\meter}). Substituting the values of $\kappa$ and $w$ in Eq. (\ref{eq:expression_interfacial_thickness}) yields an interface thickness of $\ell=1$ \si{\micro\meter} ($0.1R$), which is in excellent agreement with the simulation result shown in Fig. \ref{fig:SingleVoidConfig}. 

\bibliographystyle{elsarticle-harv}
\bibliography{library} 
\end{document}